\documentclass[twocolumn, tighten]{aastex63}

\newcommand\aastex{AAS\TeX}
\newcommand\latex{La\TeX}

\usepackage{url}
\usepackage{hyperref}
\hypersetup{colorlinks,linkcolor=red,citecolor=blue,urlcolor=blue}

\submitjournal{ApJ}

\shorttitle{Two populations in the thin disk}

\begin{document}

\title{Two sequences in the age-metallicity relation as seen from [C/N] abundances in APOGEE}

\correspondingauthor{P. Jofr\'e}
\email{paula.jofre@mail.udp.cl}

\author{Paula Jofr\'e}
\affiliation{N\'ucleo de Astronom\'ia, Facultad de Ingenier\'ia y Ciencias \\
Universidad Diego Portales,  Ej\'ercito 441, Santiago, Chile}





\begin{abstract}
The age-metallicity relation is fundamental to study the formation and evolution of the disk.  Observations have shown that this relation has a large scatter { which can not be explained by observational errors only}. That scatter is { hence attributed to the} effects of radial migration { in which stars tracing different chemical evolution histories in the disk get mixed. }
However, the recent study of \cite{Nissen_2020A&A...640A..81N}, using high precision observational data of solar type stars,   { found two relatively tight age-metallicity relations. One sequence of older and metal-richer stars probably traces the chemical enrichment history of the inner disk while the other sequence of younger and metal-poorer stars the chemical enrichment history of the outer disk. If uncertainties in age measurements increase,  these sequences mix explaining the scatter of the one relation observed in other studies. This work follows up on these results, by analysing an independent sample of red clump giants observed by APOGEE. Since ages for red giants are significantly more uncertain,   the [C/N] ratios are considered as a proxy for age.  This larger dataset is  used to investigate these relations at different Galactic radii, finding that these distinct sequences exist only in the solar neighbourhood. The APOGEE dataset is further used to explore different abundance and kinematical planes to shed light on the nature of these populations.  }
\end{abstract}



\section{Introduction} \label{sec:intro}

The age-metallicity relation (AMR) in the solar neighbourhood has been  a fundamental diagram to understand how the Milky Way disk has formed. Considering that stars pollute star forming regions with metals after exploding, younger stars need to be more metal-rich than older stars. Hence, in a closed-box scenario, it is expected that the AMR is tight, with a trend of metallicity increasing with time.  However, even early works on the AMR, either using smaller spectroscopic samples \citep{1993A&A...275..101E} or larger photometric samples \citep{2011A&A...530A.138C}, have shown the contrary: the relation has a scatter much larger than the uncertainties in determined metallicities or ages.  

Explaining the scatter of the AMR has used the fact that stars migrate and that chemical enrichment is not constant across the disk \citep[e.g.][]{2018MNRAS.475.5487S}. Therefore, the variety of ages and metallicities with apparently no relation in the solar neighbourhood can be attributed to stars tracing the chemical enrichment history from different birth places \citep{2010ApJ...722..112M}. Recently, \cite{Feuillet_2019MNRAS.489.1742F} has attempted to unveil the true structure of the AMR  in an extended region of the disk  taking advantage of large spectroscopic datasets.  By calculating means in the age distribution at a given metallicity, they found that the relation is not flat, but has a turn-around at solar metallicities, i.e., that metal-poor and metal-rich stars are older than solar metallicity stars. This can be interpreted as the effect of stellar migration, that is,  older stars born in inner regions of the Galaxy have moved outwards \citep[e.g.][]{Miglio_2021A&A...645A..85M}.


Yet, quantifying the importance of migration and the net difference in chemical enrichment rates in the disk have remained an open question. The recent work of  \cite{Johnson_2021arXiv210309838J} used the observed AMR of  \cite{Feuillet_2019MNRAS.489.1742F}  to constrain not only the amount of migration of the fossil stars, but also the migration of the intermediate-mass stars which will become white dwarfs and then explode as Supernova Type Ia (SNIa). Since the latter are long lived, they are able to explote far from their birthplaces. This enriches a different region of the Galaxy with metals.  A precise AMR is thus of paramount importance to constrain the different processes that affect the disk formation and structure.

If the precision of measured ages and metallicities improve, we could unveil the details of the AMR and so the formation of the disk.  \cite{Nissen_2018A&ARv..26....6N} extensively illustrated the power of using high-precision abundances of stars in several astrophysical applications. In particular, using high precision solar-twins  abundances it has been possible to study tight relationships between chemical abundance ratios and ages,{ such as [Y/Mg] and [Ba/Mg] \citep[e.g.][]{Nissen_2015A&A...579A..52N, Spina_2018MNRAS.474.2580S, Jofre_2020A&A...633L...9J}. These relations are attributed to a strong dependency of chemical enrichment with time.  These abundances have been dubbed as ``chemical clocks'' and offer interesting perspectives of using certain abundance ratios as an alternative for stellar ages. 

With the intention of testing the applicability of these chemical clocks, which have been shown to vary with metallicity \citep{delgado_2019A&A...624A..78D} and Galactic region since the star formation efficiency is not constant across the Galaxy} \citep[e.g.][]{Casali_2020A&A...639A.127C, 2021arXiv210314692C}, \citet[hereafter N20]{Nissen_2020A&A...640A..81N} extended the metallicity range of solar{-{type stars from the solar-twin regime ($-0.1 < \mathrm{[Fe/H}] < 0.1$) to $-0.3 < \mathrm{[Fe/H]} < 0.3$}. To their surprise, they found two notable separated sequences in the AMR of their sample.   The stars from both relations showed different chemical abundance ratios of a few other elements, as well as the overall kinematic distributions.  They concluded that if the AMR  the solar neighbourhood is truly composed by two sequences, it should be seen in larger samples as well.  They stressed, however, that such study is not straightforward due to the typical large uncertainties in ages of stars very different to the Sun. 

Fortunately, there is a way to avoid using ages and still test N20's  results in larger datasets. If ages do not reach the needed precision, we need to consider an alternative proxy for age that is precise. \cite{2015MNRAS.453.1855M} demonstrated the power of using [C/N] abundances of red giants as proxy for ages to show how the thick disk and the thin disk had very different evolutionary histories.  { This abundance ratio differs from the chemical clocks since it does not follow a chemical enrichment rate but is the product of processes happening inside red giant stars. }  \cite{2015MNRAS.453.1855M}  revived the fundamental principle that red giants change their C/N atmospheric abundance ratio after  experiencing their first dredge-up because they bring new material synthesised in their cores towards the atmosphere. This is a consequence of the CNO cycle for hydrogen burning, and applied it to Galactic archaeology. Since the change in the C/N ratio largely depends on the stellar mass, a distribution of [C/N] abundances of a large sample of red giants of similar evolutionary stages can be related to their masses, hence their ages.  


APOGEE has very precise atmospheric parameters and stellar abundances, in particular, metallicities and [C/N] abundance ratios,  for a very large number of red giants distributed in the entire Galactic disk \citep{Jonsson_2020AJ....160..120J}. 
Indeed, \cite{Hasselquist_2019ApJ...871..181H} studied the [C/N]-[Fe/H] relation in the disk as an alternative for the AMR.  They used large samples of stars in APOGEE, spanning a wide range in parameter space and abundances. That allowed them to include the thin and the thick disk in order to study the difference in the [C/N]-[Fe/H] relations and distributions at different Galactic radii and heights.  They commented that the solar neighbourhood had a larger scatter in the [C/N]-[Fe/H] relation than the majority of the Galactic regions, but no two separate sequences were identified. 

The present work attempts to focus the discussion in a possible dual AMR in the solar neighbourhood by selecting from APOGEE only stars that { have a restricted range} parameter space to avoid scatter in the [C/N]-age relation as much as possible. The main motivation is to focus on the Galactic plane and stars close to solar metallicities, following N20's results.    Such stars are then further studied taking advantage of the multidimensional information available for the APOGEE stars today, namely kinematics from Gaia and the several elemental abundances beyond $\alpha$, C and N. With this focused analysis two separate [C/N] - [Fe/H] relations are found, sharing similar properties to those of N20.

\section{Data and Methods}\label{data}

Two valued added catalogues from the 16th Data Release of SDSS \citep{2020ApJS..249....3A} were considered.  The first one is the {\tt APOGEE Red-Clump (RC) Catalogue}\footnote{\url{https://www.sdss.org/dr16/data\_access/value-added-catalogs/?vac\_id=apogee-red-clump-(rc)-catalog}} which includes the identification of RC stars using spectrophotometric data applied then to APOGEE. It has a 95\% certainty of the star being a RC star. Details of this selection can be found in \citet[]{Bovy_2014ApJ...790..127B}. The second catalogue is the {\tt astroNN catalog of abundances, distances, and ages for APOGEE DR16 stars}\footnote{\url{https://www.sdss.org/dr16/data\_access/value-added-catalogs/?vac\_id=the-astronn-catalog-of-abundances,-distances,-and-ages-for-apogee-dr16-stars}}, which consists on applying a deep learning neural network to APOGEE spectra in order to derive, among other properties,  distances as described in \cite{Leung_2019MNRAS.489.2079L}. That method is model free, because it is based on the spectra and the parallax of stars and applied to spectra of stars with unknown parallax \citep[see][for further discussions]{Jofre_2015MNRAS.453.1428J, Jofre_2017MNRAS.472.2517J}. The catalogue also includes ages and dynamical and kinematical properties such as actions, eccentricities and velocities, following the description of \cite{Mackereth_2019MNRAS.489..176M}. Both samples cross-matched gives approximately 40,000 RC stars with abundances, distances, ages and dynamical parameters. 

The atmospheric parameters and stellar abundances considered  are those of APOGEE DR16, that is,  from the APSCAP pipeline \citep{Garcia_2016AJ....151..144G}. They are based on fitting synthetic spectra computed considering LTE and 1D  to the observed spectra using the FERRE code \citep{2006ApJ...636..804A}.  The spectra from which abundances are derived are in the infrared and have resolution of about 20,000. Spectra have normally a signal-to-noise of 100, which is sufficient to have abundance precisions normally below 0.05 dex \citep{Jonsson_2020AJ....160..120J}. Extended discussion on APOGEE accuracy and precision in context with other optical spectroscopic surveys can be also found in \cite{Jofre_2019ARA&A..57..571J}. 

Further cuts were applied on the data. As a quality cut only stars with spectral parameters determined with confidence were considered. This means to take only stars with parameter {\tt APSCAPFLAG = 0}.  
In addition,  only stars in metallicity range of $-0.35 < \mathrm{[Fe/H]} < 0.35$ were selected with the intention to have a similar metallicity range to N20. In order to remove thick disk (high-$\alpha$) stars from the sample,  further chemical and spacial cuts were applied by requesting  $[\alpha/\mathrm{M}] < 0.1$, Galactic height $|\mathrm{z}| < 0.8$ kpc, and a positive parallax measured with better confidence than 20\%.  All these cuts reduces the sample to about 18,000 stars. { The atmospheric parameters of the sample have a distribution of mean temperature of 4800~K and a standard deviation of 130 K and mean surface gravity of 2.44 and standard deviation of 0.1~dex. Abundance ratio precisions have mean and standard deviation indicated in Table~\ref{tab:xfe_errors}. }

\begin{table}[t]
\caption{Mean and standard deviation of the error distributions in the abundance ratios considered in this work.}
\begin{center}
\begin{tabular}{c|cc}
\hline
$\sigma$[X/Fe] & mean & standard deviation  \\
\hline
C & 0.012 & 0.004 \\
N & 0.018 & 0.005 \\
O & 0.015 & 0.005 \\
Mg & 0.011 & 0.002 \\
Al & 0.02 & 0.005 \\
Si & 0.011 &  0.002 \\ 
Ca & 0.013 & 0.004 \\
Ti & 0.018 & 0.0058 \\
Cr & 0.035 &  0.010 \\
Mn & 0.016 & 0.005 \\
Fe & 0.008& 0.0002 \\
Ni & 0.013 & 0.007 \\
\hline

\end{tabular}
\end{center}
\label{tab:xfe_errors}
\end{table}%

\subsection{[C/N] as a proxy for age}
\cite{2015MNRAS.453.1855M} took upon the work of \cite{1965ApJ...142.1447I} to demonstrate that indeed [C/N] abundances could be used to study the mass, hence age, distribution of red giant populations. Since then,  there has been rich literature regarding the relation of [C/N] abundances and ages for red giant stars. Some works derive empirical relations of measured [C/N] abundances from the spectra with independent measurements of ages from e.g. asteroseismology \citep{2016MNRAS.456.3655M} or open clusters \citep{2019A&A...629A..62C}. Other works derive ages using {\it The Cannon} directly to the spectra \citep{2016ApJ...823..114N}. The latter, however, indirectly uses [C/N] as important input for ages \citep[see also discussions in ][]{2019MNRAS.484..294D}. 

\begin{figure}[t]
\centering
\includegraphics[scale=0.5]{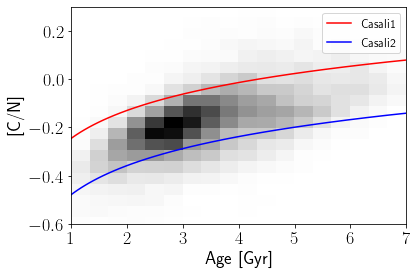}
\caption{Correlation of [C/N] with age for the sampled stars. For reference, the empirical relation of age and [C/N] determined using open clusters by \cite{2019A&A...629A..62C} is shown with red and blue lines, { and correspond to the relations $ \log{\mathrm{Age (yr)}} = 10.64 + 2.61 \mathrm{[C/N]}$ and $\log{\mathrm{Age (yr)}} = 11.20 + 2.51 \mathrm{ [C/N]}$, respectively. }  }
\label{fig:casali}
\end{figure}

\begin{figure*}[t]
\centering
\includegraphics[scale=0.5]{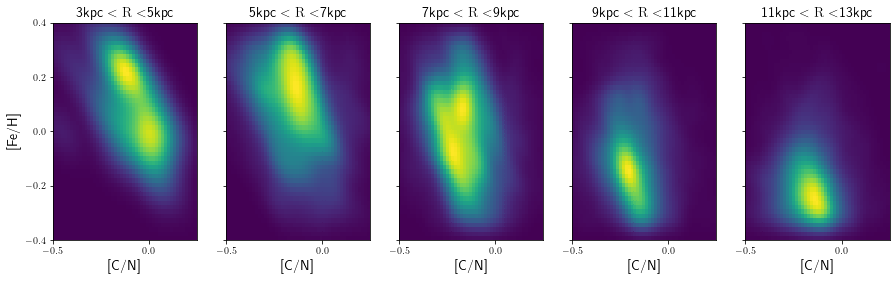}
\caption{Distribution of [C/N] and and [Fe/H] at different galactocentric radii. }
\label{fig:cn_feh_maps_r}
\end{figure*}

Figure~\ref{fig:casali} shows the relation of [C/N] abundance ratios and ages of the stars used here. The ages considered for this plot are those from {\tt AstroNN}, which were derived by \cite{Mackereth_2019MNRAS.489..176M}. They discuss the dependency of [C/N] in the derived ages, therefore this relation is expected. For guiding the eye, the empirical relationship determined by \cite{2019A&A...629A..62C} is plotted with lines. They used  [C/N] abundances of red giants in open clusters observed with the APOGEE and the Gaia-ESO \citep{2012Msngr.147...25G, 2013Msngr.154...47R} surveys and independent ages derived from isochrone fitting to the CMD of the clusters to find an empirical relation.    Their results have a generous uncertainty in the zero point and slope of the relation, allowing for a range of options between [C/N] abundances and ages.  Here, the relations
$ \log{\mathrm{Age (yr)}} = 10.64 + 2.61 \mathrm{[C/N]}$ and $\log{\mathrm{Age (yr)}} = 11.20 + 2.51 \mathrm{ [C/N]}$
are plotted with red and blue, respectively. 

The figure intends to show that the [C/N] abundances in this sample can be used as an age proxy. This is important step considering that RC stars are  avoided in some studies of [C/N] abundances with age \citep[e.g.][]{2015MNRAS.453.1855M, Hasselquist_2019ApJ...871..181H}.  RC stars might have experienced extra mixing processes while during the red giant branch through the helium flash or thermohaline mixing which might have altered the [C/N] abundances after the first dredge-up \citep{Masseron_2017MNRAS.464.3021M, Lagarde_2017A&A...601A..27L}.  The figure shows that this sample, which is restricted to a tight space in stellar parameters, still leads to a correlation with age, but transforming [C/N] abundances into ages can imply uncertain ages, especially because these mixing processes are poorly understood.

\section{Results}

\subsection{[C/N]-metallicity as a function of Galactic radius}

Figure~\ref{fig:cn_feh_maps_r} shows density maps in the [C/N]-[Fe/H] plane for stars located in different Galactocentric radii $R$.   
The panels show the [C/N]-[Fe/H] for stars located from the inner to the outer disk, in $R$ bins of 2 kpc starting at 3 kpc on the left hand panel and finishing at 13 kpc on the right hand panel. 

It is possible to notice the sark difference of  [C/N]-[Fe/H] relations between the different panels. The first panel ($3<R<5$ kpc), containing 62 stars only,  shows two separated groups of stars, with different metallicities and [C/N], which could be attributed to the overlap between the inner disk and the bulge. The second panel ($5<R<7$ kpc), containing  { 1488} stars,   shows rather one sequence of stars, relatively metal rich and high in [C/N] abundances. The stars follow a tight relation between [C/N] and [Fe/H], suggesting they  experienced a rapid chemical enrichment in the past, probably like a close-box.  The panel further shows a large number of stars in the background which are not part of this relation and show indication of a secondary sequence at lower metallicities, but there are not enough stars to confirm this.    

The third panel ($7<R<9$ kpc), containing { 7016} stars,  corresponds to the solar neighbourhood.  This region shows there are two separated groups of stars. The first one is a sequence which has similar properties to the sequence seen in the adjacent left panel, namely its stars are rather metal-rich and [C/N] enhanced, and a have steep relation. The second sequence, which could be a continuation of the background population of the previous panel, shows a tight relation between [C/N] and [Fe/H], reaching  more metal-poor stars and higher [C/N] values.  
The groups are separated, agreeing with N20's results about the two sequences in the solar neighbourhood. These sequences were not seen in \cite{Hasselquist_2019ApJ...871..181H} probably because they used stars covering a wider range in stellar parameters, which might have blurred the distributions. In fact, that same panel was commented to have a larger scatter in the [C/N]-[Fe/H] relation with respect to the rest of the panels. 

The fourth panel ($9<R<11$ kpc) contains { 7322} stars and presents one main sequence of stars rather metal-poor as well as [C/N]-poor. This suggests that the outer disk has had a slower star formation history, with  stars being rather young. There is however a tight relation between [C/N] and [Fe/H], similar to the the solar neighbourhood lower sequence.  Like in the ($5<R<7$ kpc) panel, there is significant background of stars which do not follow the relation. Here, however, they are rather metal-rich, and could be attributed to the tail of the upper sequence seen in the solar neighbourhood panel.  The last panel ($11<R<13$ kpc) contains { 1764} stars of almost only metal-poor stars with low  [C/N], suggesting these stars are probably rather young.  This is in agreement with previous studies, which have found a dominance of younger stars in the outer disk \citep{2017ApJS..232....2X, 2020ApJS..249...29H}. The relation between [C/N] and metallicities in this panel is rather loose.  

\cite{Feuillet_2019MNRAS.489.1742F} also plotted the AMR at different Galactic radii and heights. Comparing with their results for the Galactic plane, it is possible that their  turn-around in the AMR leading to a C-shape is attributed to the appearance of the second sequence, which, as extensively discussed by N20, it could produce a large scatter for typical uncertainties in stellar ages. The C-shape in the AMR appears in the outer disk panels as well, but in Fig.~\ref{fig:cn_feh_maps_r} the secondary sequence tends to disappear. This might be a selection effect, since here only RC stars are considered, while \cite{Feuillet_2019MNRAS.489.1742F} include a wider range of surface gravities hence luminosities.  

{ It is worth to comment the selection effects of these distributions due to the bias induced by selecting only metallicities that are above -0.3 dex. Because the star formation efficiency is different across the disk, the stars of same metallicity in the inner and outer parts of the disk do not necessarily reflect the same timescales and epochs of formation and hence, they can not be interpreted in a total evolutionary framework. Adding however a wider range in metallicity might induce further scatter in the C/N-age relation \citep{Das_2020MNRAS.493.5195D} and other systematics which this work is trying to avoid.  }

\begin{figure}[t]
\centering
\includegraphics[scale=0.5]{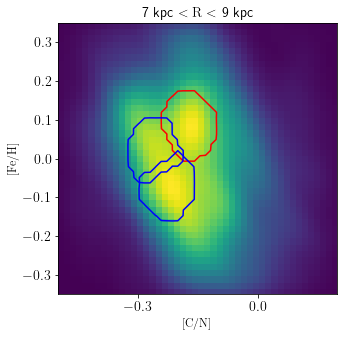}
\caption{Density map of [C/N] and [Fe/H] in the solar neighbourhood alongside with the 3 main clusters found using Shift Mean. }
\label{fig:clusters}
\end{figure}
 
\begin{figure*}[t]
\centering
\includegraphics[scale=0.35]{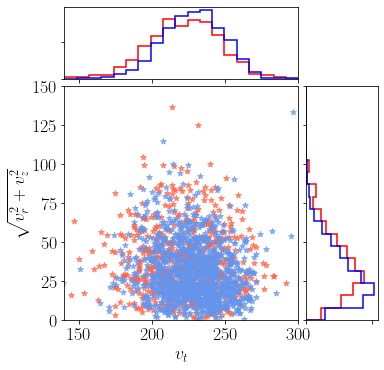}
\includegraphics[scale=0.35]{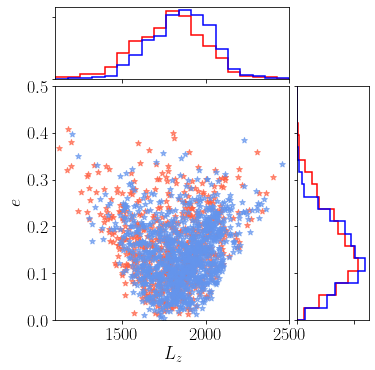}
\includegraphics[scale=0.35]{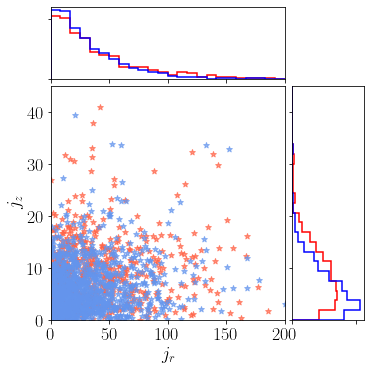}
\caption{Kinematics of  selected groups. The colors follow the classification from Fig.~\ref{fig:clusters}. }
\label{fig:kinematics}
\end{figure*}

\subsection{Two populations in the solar neighbourhood}
In order to see if the two sequences of AMRs found by N20 are also present in this larger and independent dataset,  the density map of [C/N] and [Fe/H] for the stars located in the solar neighbourhood  is shown again in Fig.~\ref{fig:clusters}. If these sequences are produced due to two different populations tracing different chemical enrichment histories, investigating more chemical patterns as well as kinematics is useful.  Therefore,  only stars located at the maxima of the distributions were selected considering the clustering algorithm Shift Mean, which is designed to find local bumps in a density estimate of data \citep[see Chapter 6.4 of][]{astroMLText}.  The data of the solar neighbourhood supports { 3 main} groups. { which are located in the main two populations. One is more metal-rich (reaching [Fe/H] of 0.3 dex) and can be considered to be older because the [C/N] ratios are higher (see e.g. Figure~\ref{fig:casali}) and the other one has a more extended range in metallicity reaching lower [Fe/H] values of -0.2 as well as a more extended range in [C/N] suggesting a larger range in ages reaching younger ages than the more metal-rich and older population. }

The groups are shown as coloured contours in Fig.~\ref{fig:clusters}. The stars from the upper population are given a red colour, while the stars from the lower population groups are given blue colours. These are studied in terms of their kinematics and other elemental abundances.  

\subsubsection{Kinematics}

Figure~\ref{fig:kinematics} shows the distributions of different dynamical quantities for the stars enclosed in the red and blue groups of Fig.~\ref{fig:clusters}.

The left-hand panels show the distributions of the stars in the classical Toomre diagram, which was also plotted by N20. While there is a large overlap between both groups, the red group has a tendency of having lower tangential velocities but slightly higher  vertical and radial velocities than the blue group.  This result agrees with N20, who also found that their red (old) group had a larger velocity dispersion and a larger rotational lag than the blue (younger) group, which is encouraging. Differences in kinematics are in any case minimal, here and in N20, making it difficult to add more about the origin of these two groups from the velocities alone.

\begin{figure*}[t]
\centering
\includegraphics[scale=0.35]{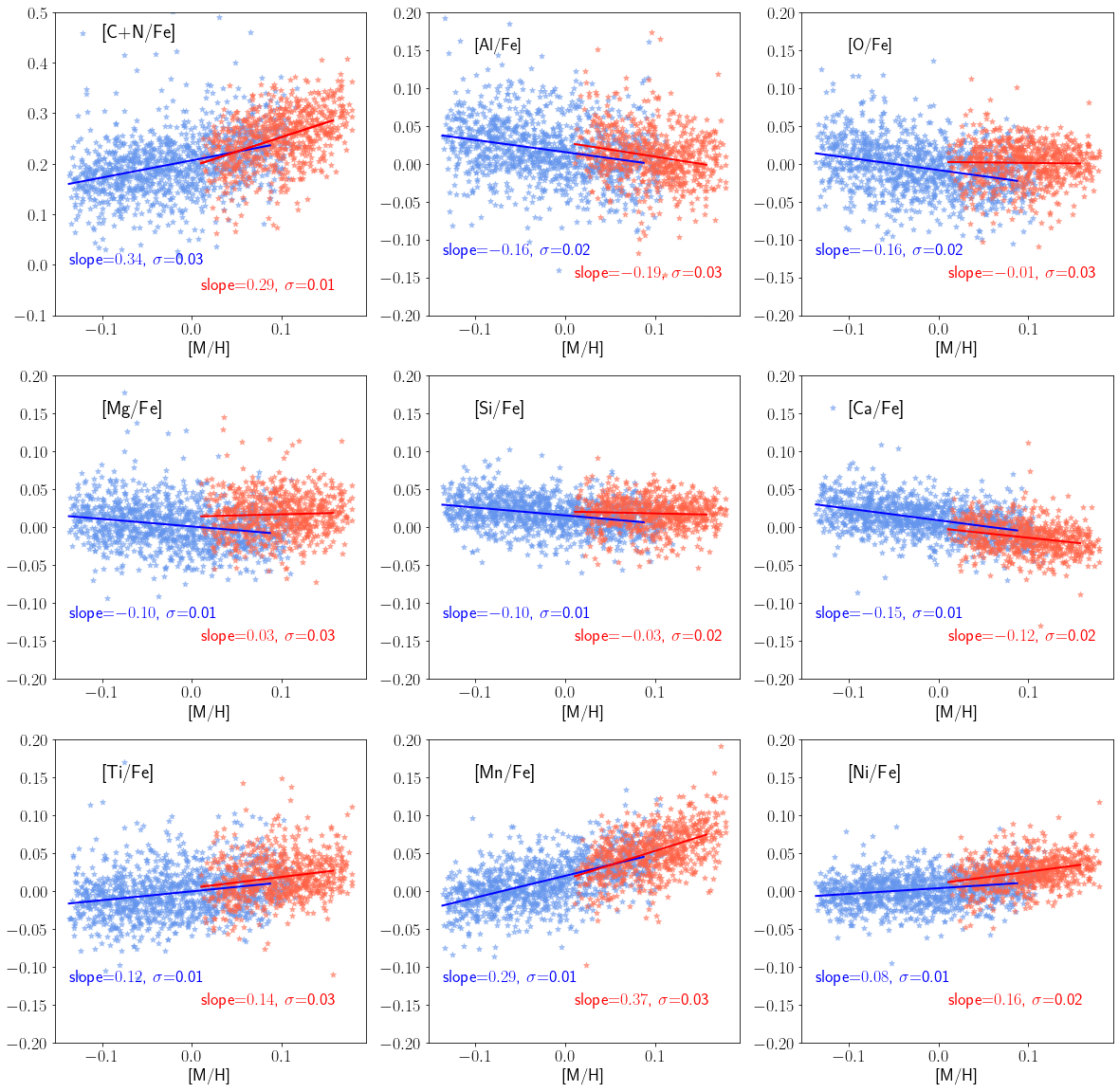}
\caption{Abundance ratios as a function of metallicity of selected groups from the [C/N]-[Fe/H] diagram. Linear regression fits to the data of the corresponding groups are displayed for guidance.  }
\label{fig:abundances1}
\end{figure*}

The middle panels plot the relation of the stellar orbits angular momenta $L_z$ and the eccentricities $e$.  As in the left-hand panel, the overlap between both groups is large, and only slight differences can be identified from the histograms. Both groups have relative circular orbits, with a range of eccentricities comparable and below $e<0.4$. The red group presents, however, a tendency of having stars slightly more eccentric at $0.3<e<0.4$ than the blue group, indicating that more stars in the red group have been kinematically heated. This makes sense since the red group has overall higher [C/N] abundance ratios than the blue group, hence the red group has stars that are generally older, and  old stars are kinematically hotter than young stars \citep[e.g.][]{2008A&A...480...91S, Mackereth_2019MNRAS.489..176M, 2020arXiv200406556S, 2021MNRAS.503.1815B}.   

 While the overlap in $L_z$ is large, the red group has overall lower angular momenta than the blue group. If the red group is associated to a sequence of stars tracing the chemical enrichment history of the inner disk, then they could have radially migrated. The `churning' effect of radial migration move stars outwards by keeping eccentricities fixed but loosing angular momenta  \citep{2020MNRAS.493.1419F}, which is consistent with the behaviour of the distributions of the red group compared to the blue group.  However,  having two distributions with different $L_z$ does not necessarily mean one population has lost angular momentum, since eccentric orbits might also be an indicator of just visitors.   In fact, the right-hand panels show the radial and vertical actions $j_r$ and $j_z$, respectively. Again, both groups heavily overlap in the scatter plot, but the distributions allow for a better inspection of possible differences. While the radial action distribution is  very similar for both groups, there is a difference in the vertical actions $j_z$.   

The difference in $L_z$ and $j_z$ between the red and the blue sequences might thus be interpreted as stars being born in different regions. More specifically, the red sequence might show stars formed at smaller Galactic radii that due to their higher eccentric orbits compared to the blue sequence reach the solar neighbourhood.

\subsubsection{Elemental abundances as a function of metallicity}

The red and the blue groups  are also studied in other abundance planes, which are shown in Fig.~\ref{fig:abundances1}. 
In each panel a scatter plot of stars coloured by red and blue for different abundance ratios  as a function of metallicity is shown.  Linear regression fits to the corresponding distributions of the respective groups as a function of metallicity have been performed and are indicated with solid lines of the same colour.  The slope and the error of the regressions are indicated in each panel. 

In all panels the abundance trends have a smooth transition between one and the other sequence.  For most of the elements both sequences merge into one sequence, behaving as the classical {\it low-$\alpha$} chemical sequence found in the APOGEE data \citep{Jonsson_2020AJ....160..120J}. 
The slope of the regression fits of the abundances as a function of metallicity, however, can differ between the red and blue group for some cases, such as [C+N/Fe], [O/Fe], [Mg/Fe], { [Si/Fe] } and [Ni/Fe]. 

As discussed in \cite{2015MNRAS.453.1855M}, while the [C/N] abundance changes in giants after they experience dredge-up, the total amount of C and N stays the same, reflecting the composition of the birth cloud, which, like many other chemical abundances, will change as stars die \citep[see also Fig. 2 of][]{2016MNRAS.456.3655M}. 
Hence, [C+N/Fe] ratios  have been used to study the differences of star formation histories in the disk \citep{2015MNRAS.453.1855M, Hasselquist_2019ApJ...871..181H} as well as metal-poor stars which might have form in dwarf galaxies and accreted later on in the Milky Way \citep{Hawkins_2015MNRAS.453..758H, Das_2020MNRAS.493.5195D, Horta_2021MNRAS.500.1385H}. Around solar metallicities, C is made by both, He-burning in the core of stars as well as in AGB stars, while N is made mostly in AGB stars \citep{Kobayashi_2020ApJ...900..179K}, therefore [C+N/Fe] traces the contribution of AGB stars.  The fact that the red group has overall higher [C+N/Fe] abundances than the blue group might be attributed to the difference in metallicity, since both C and N have a metallicity dependency, and the red group has mostly metal-rich stars while the blue group has a wider range in metallicities.  { It is noted that the} [C+N/Fe] are above zero even for the blue sequence. This might be related to a systematic uncertainty in the N abundances of APOGEE, { which obtain a value of 0.2  for the solar N abundance } \citep{Jonsson_2020AJ....160..120J}.  Indeed, the recent study of nitrogen abundances in the Sun of \cite{2020A&A...636A.120A} points towards a systematic difference between N measured from molecular features and atomic features. { It is not the aim of this paper to correct by systematic effects but to illustrate the trends in both populations and look for differences. This is another argument of why this work does not attempt to relate [C/N] directly with an age value.}

Oxygen and magnesium are $\alpha-$capture elements that are produced inside massive stars and are ejected into the interstellar medium via core-collapse supernova (SNII). Both oxygen and magnesium belong to the few primary elements, e.g., the yields are not affected by the metal content of the progenitor star. Iron, on the other hand, is mostly produced by thermonuclear supernova (SNIa), whose progenitors are lower mass stars. Therefore there is a time delay in the enrichment of oxygen (or magnesium) and iron \citep{1986A&A...154..279M}. There is extensive literature in using in particular [O/Fe]-[Fe/H] planes to constrain chemical evolution models, both of the Milky Way \citep{Johnson_2021arXiv210309838J} and other galaxies \citep{2019A&ARv..27....3M}.  The mass of the progenitor galaxy as well as the star formation history can be addressed from the relation between [$\alpha$/Fe] and [Fe/H] for a given stellar population \citep{1979ApJ...229.1046T}. The [$\alpha$/Fe]-[Fe/H] -- hereafter Tinsley-Wallerstein (TW)\footnote{This is motivated after discussions among astronomers on Social Media in 2020 about calling this important diagram after Wallerstein and Tinsley who used and explained it. Many fundamental diagrams are named after scientists that have first explained.} -- diagram on the top right hand panel of Fig.~\ref{fig:abundances1} shows how the blue group has lower [O/Fe] ratios than the red group for the same metallicity. It further shows that while the red stars  seem to have reached a plateau at solar metallicities, the blue stars still are in the decreasing part of the TW diagram, suggesting that star formation efficiency is lower for the blue group.  A similar behaviour is seen for Mg, shown in the left-hand side panel of the middle row of Fig.~\ref{fig:abundances1}.   It is interesting to note that these abundance planes in the APOGEE data show a turn-around in the abundances around solar metallicities \citep{Jonsson_2020AJ....160..120J}, which might also be seen in the [Mg/Fe] - [Fe/H] trends with optical spectroscopy \citep{Adibekyan_2012A&A...545A..32A}.

{ Silicon is another $\alpha$-capture element which presents a slight change of slope in the regression fits between the blue and the red population. This element is believed to be produced by both, SNII as well as SNIa, hence it does not correlate directly with Mg which is mostly produced by SNII. Furthermore, the production mechanism in SNII is different to Mg which translates to a dependency on the progenitor's mass \citep{2019ApJ...883...34B}.  In the populations of this study, the Si abundance is a combination of the many SNII of lower mass, in addition to SNIa. The fact that  the red and the blue groups have a slight difference can be an effect of the different metallicities between the groups (SNIa contribution) as well as age (SNII contribution).  }


Nickel is an iron-peak element which is synthesised in SNIa as well as inside massive stars and expelled into the interstellar medium via explosions in essentially the same way as Fe. It is therefore expected that [Ni/Fe] has an overall flat and solar value for an extended range in metallicity.  But as recently discussed by \cite{Kobayashi_2020ApJ...895..138K}, the large variety of possible progenitors of SNIe, makes it hard to reproduce and interpret the observed [Ni/Fe] abundance ratios, especially at high metallicities \citep[see also][for the effect of yields of different SNIe prescriptions]{2021MNRAS.503.3216P}.  Since the yields differ between the different white dwarfs progenitors of the supernovae, the elemental abundance ratios will be affected when changing the contribution of  such progenitors.  While the metal production of SNIa is independent of metallicity, it production rate can be affected by metallicity, since the lifetime of the secondary star of the binary depends on the progenitor metallicity.  

The red sequence  shows a steeper trend of [Ni/Fe] with metallicity than the blue sequence. This could be attributed to a different metallicity effect for the progenitor (hence timescale) of the SNIe, affecting in a different way the [Ni/Fe] abundance ratios. It would be the difference of environments leading to different contributions form SNIa subclasses \citep{2021MNRAS.503.3216P}.  The slight increase at metallicities above solar has been seen in the literature \citep{Adibekyan_2012A&A...545A..32A, Bensby_2014A&A...562A..71B} as well as when considering the bulk of the APOGEE data \citep{Jonsson_2020AJ....160..120J}. Considering the break in the slope between both sequences, it is interesting to associate this increase to an overlap between two stellar populations coexisting in the solar neighbourhood which trace different star formation histories.

\begin{figure*}[t]
\centering
\includegraphics[scale=0.35]{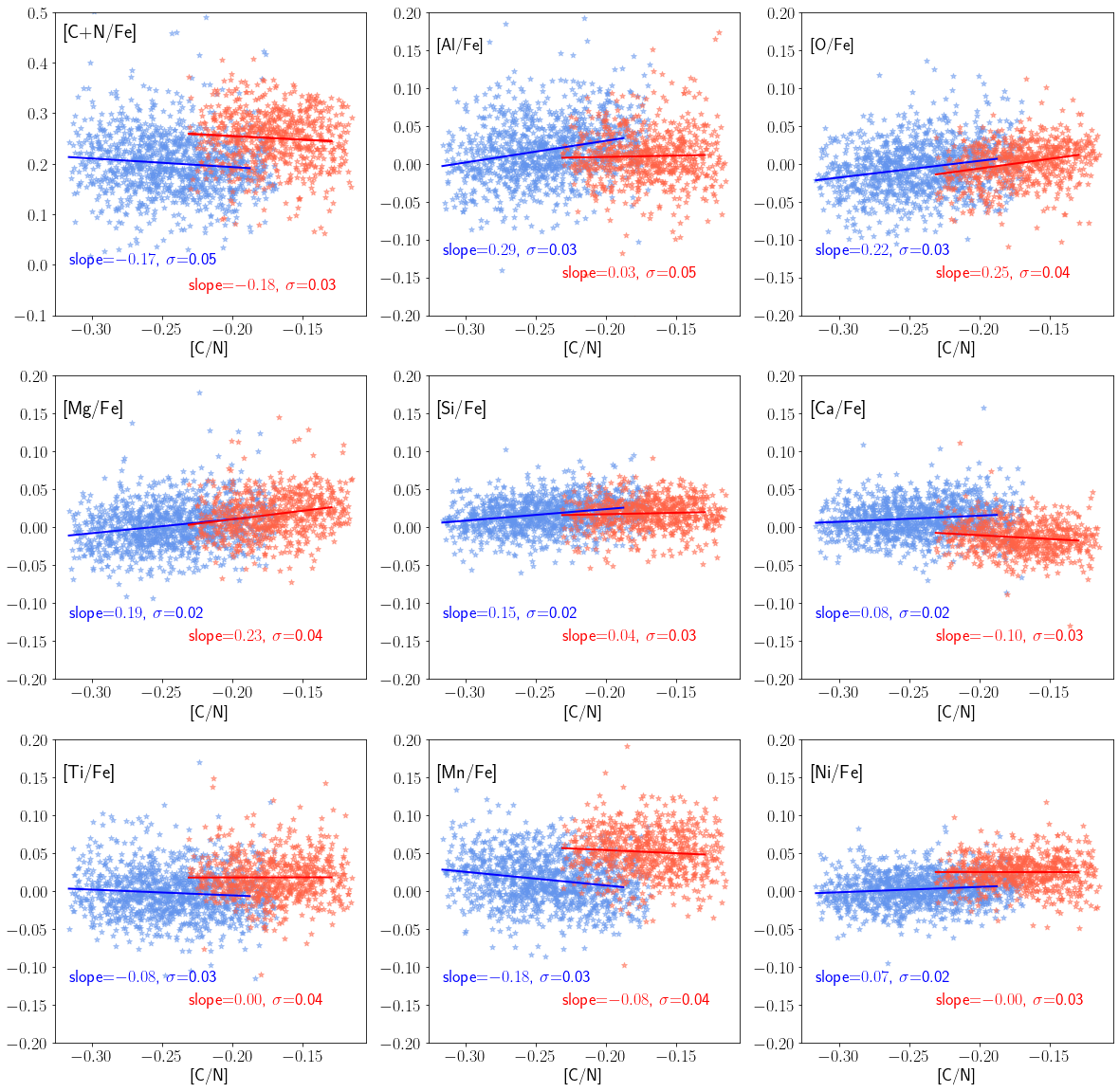}
\caption{Abundance ratios as a function of [C/N] for the same sample as in Figure~\ref{fig:abundances1}. }
\label{fig:abundances}
\end{figure*}

\subsubsection{Elemental Abundances as a function of [C/N]}

Figure~\ref{fig:abundances} presents the same abundance planes as in Fig.~\ref{fig:abundances1} but here as a function of [C/N]. This allows for an alternative to study the temporal evolution of the elements in each sequence, and thus chemical enrichment rates.  Following the previous section, linear regression fits to the data have been performed to help interpreting the results. When studying the abundance ratios in the [C/N] plane  most of the merged  sequences seen in Fig.~\ref{fig:abundances1} break down in two separate sequences. 

Interestingly, the $\alpha$-capture elements O and Mg, which showed a sequence in the TW diagram of Fig.~\ref{fig:abundances1} with different slopes, merge into one sequence in Fig.~\ref{fig:abundances}. This might be due to the primary nature of this element, which is produced by massive stars regardless of the initial metallicity of the stars.  { Silicon, on the other hand, has different slopes for the blue and the red sequence, providing evidence of the SNIa production of this element, in addition to SNII. }
 Tight [O/Fe] or [Mg/Fe] with age relations have been discussed in the literature \citep{2018MNRAS.475.5487S, delgado_2019A&A...624A..78D, Haywood_2019A&A...625A.105H, Hayden_2020arXiv201113745H}. That the relation between [O/Fe] or [Mg/Fe] with [C/N] is tight reinforces the postulation of using [C/N] as a proxy for precise ages in RC stars. 


Abundances [C+N/Fe] present a large scatter, with a slight negative trend as a function of [C/N] for both groups.  Since at these metallicities AGBs produce C and N, the abundance ratio is expected to increase with time, hence showing a negative trend with [C/N]. Both sequences have an offset, suggesting that the red stars have had more pollution from AGB stars than the blue sequence at a given time. That could happen if star formation is fast, reaching higher metallicities at a given time. The higher metallicities serve as seed for higher C and N production in AGB stars.  

While no significant offset between the [Al/Fe] mean abundances of the blue and the red groups is found, the overall trends show opposite slopes, with the blue sequence increasing with [C/N] while the red sequence decreasing with [C/N].  The regression fits are however uncertain, since the abundances have high scatter.  Aluminum is an odd-z element which is produced inside stars and is expelled into the interstellar medium via core-collapse supernovae. More specifically, its production yields from nuclear reactions which use Ne and Na as seeds.  While these are also produced generally inside stars, higher metallicity stars have a larger Al production because some seeds are already available \citep{Kobayashi_2020ApJ...900..179K}. This metallicity dependency shows as a stronger effect in the different sequences.   

Calcium shows a stark discontinuity between the red and the blue sequence in the [Ca/Fe]-[C/N] plane, breaking down from the TW diagram shown in Fig.~\ref{fig:abundances1}.  The abundances follow a tight relation that have opposite trends, in which the blue sequence is positive and the red sequence is negative. This is very similar to the case of [Al/Fe], which can be understood from the same arguments. Calcium is an $\alpha-$capture element produced in massive stars, but is a secondary element whose yields depend on metallicity. Abundances of Ca are in general well-measured, there are many clean lines with accurate atomic data for calcium in the APOGEE spectra, as well as in optical spectra \citep{Jofre_2019ARA&A..57..571J}, therefore the trends here are more accurate than in the case of aluminum. 

 The bottom panels of Fig.~\ref{fig:abundances} show the abundances of titanium, manganese and nickel.  All of the abundance ratios show similar behaviour in their differences between  the blue and the red stars, in which the trends as a function of [C/N] have similar direction but a systematic difference with the red stars being overall more enhanced.  Both [Ti/Fe] and [Mn/Fe] have large scatter in their abundances, but [Ni/Fe] is tight. Titanium is a difficult element since it behaves like an $\alpha-$capture element and therefore commonly associated to that family, even if its production mechanism is not like the rest of the $\alpha-$capture elements. Modern theoretical yields of Ti do not match the observations \citep{Kobayashi_2020ApJ...900..179K}, making it difficult to interpret why both sequences have an offset in the [Ti/Fe] - [C/N] plane. This offset is also seen in N20. 
 
 Mn and Ni on the other hand are produced in SNIa, and given the variety of progenitors for SNIa, including the binary companion of the exploding white dwarfs, both elements have a dependency in metallicity \citep{Kobayashi_2020ApJ...895..138K} but also in the explosion mechanism \citep{2021MNRAS.503.3216P}.  Manganese, while difficult to measure because of the strong hyperfine structure splitting in their lines, is one of the best suited elements to test chemical evolution caused by SNIa. The stark differences seen in both sequences in the [Mn/Fe]- [C/N] plane hint towards two populations having formed in different environments, leading to different types of SNIe. Two stars of the same [C/N] (e.g. age) have quite different [Mn/Fe]. This is reassured by the [Ni/Fe] - [C/N] panel. The differences in [Ni/Fe] are smaller for coeval stars as in the case of manganese, but still significant given the overall scatter in [Ni/Fe] abundances is much tighter.

It is worth to comment on the abundance planes that are in tension with N20, namely [O/Fe], [Mg/Fe], [Al/Fe] and [Ca/Fe].  
N20 found that [O/Fe] had a large difference between sequences, while Mg a change of trend. In addition, the red sequences of N20 for [Al/Fe] and [Ca/Fe] were positive and steeper compared to the blue sequences. Here we obtain negative trends.   There are some possible explanations that can help understanding this difference. First,  there is a selection effect in which an [$\alpha/$M] cut has been applied to select the thin disk only. This means that here it is not possible to reach high values of Mg, O and Ca of N20. Second, the abundances might have significant offsets between both samples, since they are determined from very different methods and spectral signatures (fitting molecules in the case of APOGEE and determining equivalent widths of atomic lines in the case of N20). This can lead to large differences in final abundances \citep{2017ASInC..14...37J}. Third, the ages of the stars from  the N20 red sequence can be up to 10 Gyr, whereas here the sequence might reach 6 Gyr at most (see Fig.~\ref{fig:casali}). Furthermore, the [C/N] range of the red sequence ($\sim 0.1$ dex) implies a range in age of few Gyr, which makes the ``age-trend" for the red group based on the [C/N] abundances not directly comparable with N20's age sequences.

 \subsection{Uncertainties}
 
 \subsubsection{3D and non-LTE effects}
 
While the different trends might be interpreted as an overlap of populations tracing different star formation histories, it is important to consider possible systematics caused by 3D and non-LTE effects.  For example,  \cite{2019A&A...630A.104A} discussed how O abundances are affected by 3D and non-LTE effects at high metallicities. When an accurate prescription of the O triplet lines is considered for the determination of O, the apparent plateau of [O/Fe] observed at high metallicities is not seen. APOGEE results are obtained from molecular features, under the prescription of 1D and LTE, and present a plateau.
The separation between the sequences in the [C/N]-[Fe/H] relation occurs at that metallicity range, and might be responsible for part of the differences found in the TW diagram in Fig.~\ref{fig:abundances1} for both sequences. 

Other elements might also be affected by this. The recent work of \cite{2020A&A...642A..62A} illustrates the differences between LTE and non-LTE abundances for dwarfs and giants in the GALAH survey \citep{Buder_2020arXiv201102505B}. While it is not possible to directly compare the effects of giants in optical and IR spectra, one can expect differences of the same order of magnitude. In that work, abundances that have a metallicity dependency on this effect at solar metallicities  in giants are C, Mg, Al and Mn.  Indeed, \cite{Jonsson_2020AJ....160..120J} attributed some of the Mg features at solar and super-solar metallicities in the APOGEE data due to this effect.    As commented in  \cite{Jonsson_2020AJ....160..120J}, the next data releases of APOGEE will include non-LTE corrections. It will be interesting to see if the breaks found between red and blue sequence remain after these effects have been considered.

 \subsubsection{Contaminations and biases in RC stars}\label{sect:biases}
 
  The results presented here are based on the assumption that [C/N] correlates with age. In order to minimise systematic effects in this correlation, only the RC stars have been considered. This allows to have a sample of well-determined distances \citep{Leung_2019MNRAS.489.2079L} as well as stars with very similar stellar parameters. The later is necessary to relate differences in abundances with some astrophysical reason and not because of systematic uncertainties in the spectral analysis \citep{Nissen_2018A&ARv..26....6N}. 
  
  Still, restricting the parameter space does not ensure that [C/N] might be affected by other inner processes inside red giants, for example rotation \citep{2015A&A...583A..87S} or theromohaline mixing \citep{Lagarde_2017A&A...601A..27L}.  Other fundamental problems in our understanding on stellar evolution theory still imply uncertainties in ages which makes a direct estimate of [C/N] and age in red giants difficult, such as possible dependencies of metallicity with the mixing length theory \citep{2017ApJ...840...17T} or simply the stellar evolution code employed \citep{2020arXiv201207957T, 2020A&A...635A.164S}. Binary evolution and mass transfer can also  induce scatter in the [C/N]-age relation. If stars transfer mass from a binary, the [C/N] could be affected, hence can not be directly used as a mass (or age) proxy. Some such stars might not show signatures in the spectra that could hint towards a binary evolution, because the binaries could have merged and be now a single star  \citep{Jofre_2016A&A...595A..60J, 2018MNRAS.473.2984I}. Distinguishing a star that is younger from one that has experience mass transfer from the [C/N] abundances is still not obvious \citep{2019MNRAS.487.4343H}.

Using RC stars  can further induce biases in the [C/N]-age correlation. Actually, \cite{2015MNRAS.453.1855M} avoided using RC stars in their discussions, because  from their spectral parameters alone it is difficult to disentangle them with lower red giants branch (RGB) stars  which are at an earlier evolutionary phase \citep[see also][]{2017A&A...597L...3M}.  \cite{Hasselquist_2019ApJ...871..181H} also avoided the parameter space of the RC in their [C/N]-[Fe/H] relations. Using RC stars can be a problem because low mass stars experience further mixing and mass loss while at the tip of the RGB, e.g. before settling in the red clump. Stars of higher mass do not experience this since the phase at the tip of the RGB is very short. This issue was investigated in \cite{2016MNRAS.456.3655M}, who studied an empirical relation between ages and [C/N] abundances using the first APOCASK catalogue \citep{2014ApJS..215...19P}.  That sample has stars with APOGEE spectra (determining same C and N abundances as here) but also with  astroseismic observations from {\it Kepler}. The latter allows to know from the power spectrum of such observations the evolutionary phase of the stars.   \cite{2016MNRAS.456.3655M} could benefit from the seismic analysis and distinguish stars in the RC and the RGB,  deriving two different age-[C/N] relations, yet both with the comparable accuracies \citep[see also ][]{Lagarde_2017A&A...601A..27L}. They further applied the age-[C/N] relation derived for the RC in APOKASC to the entire RC catalogue of \cite{Bovy_2014ApJ...790..127B} stars and found consistent results as when applied to APOKASC only. 

\cite{Masseron_2017MNRAS.464.3021M} also studied the effect of extra mixing in RC stars by comparing nitrogen abundances of APOGEE (and APOKASC) stars with different RGB models considering extra mixing such as thermohaline mixing. While theoretical predictions hinted towards an increase in N along the RGB because of extra mixing, the observations of thin disk solar metallicity stars did not support the predictions. In fact, only observations of metal-poor stars showed an increase of N along the RGB.  They explained this by a mass effect, namely that their thin disk solar metallicity sample was in general more massive than the metal-poor sample. The higher the mass of the star, the shorter its RGB phase, hence the smaller the effects of extra mixing.  The stars used here are selected to be solar-metallicity thin disk stars, hence the effects of extra mixing altering the [C/N] abundances as proxy for age should be small. 


It is finally worth to comment that by selecting the RC in the thin disk at solar metallicities there is a bias induced towards young stars.  \cite{Bovy_2014ApJ...790..127B} discussed how these biases are increased in their RC catalog because of the selection cuts, stressing that the RC does not randomly sample the underlying age distribution of stars, but is instead the distribution is skewed toward younger  ages.  
Further noise can be due to contamination in the RC selection, since the  RC catalog has a 95\% purity \citep{Bovy_2014ApJ...790..127B}. These outliers, however, while present, still allow to see that there is an overall trend of [C/N] and [Fe/H] which results on different chemical enrichment histories in the disk, which is the main focus of this work.

\section{Discussion}

\subsection{The evolutionary history of the Galactic disk} 
 
 A interesting prospect of taking the chemical abundances for studying the evolution of stellar populations is interpreting phylogenetic trees \citep{Jofre_2017MNRAS.467.1140J, Jackson_2021MNRAS.502...32J}. As explained in these papers, using phylogenetic trees in this context is possible because there is heredity  between stars through the chemical abundances, in addition to descent with modification, i.e.,  each stellar generation is more metal-rich than the previous one, hence modifying the chemistry of the ISM. Because understanding and quantifying these two processes is at the core of phylogenetic studies,  trees are a suitable tool that can be applied in Galactic chemical evolution studies. 
 
 \cite{Jackson_2021MNRAS.502...32J} used abundances of solar twins that are very similar to those used by N20 to build such a tree. It showed two main stellar populations (branches) which were attributed to the thin and the thick disk. \cite{Jackson_2021MNRAS.502...32J} found that the root of the thin disk branch contained a group of stars whose branching pattern  was poorly supported. This group  had stars of very similar abundances and ages. Their kinematics suggested their origin spreads the disk. The authors concluded this population could  be part of the old thin disk and perhaps product of a star formation burst.  The blue sequence in fact has two bumps (see Fig.~\ref{fig:clusters}), in particular one at lower metallicities and higher [C/N]. It could be that this bump is the possible product of the star formation burst seen in the phylogenetic tree of \cite{Jackson_2021MNRAS.502...32J}.  { This bump could be a also a simply selection effect, therefore concluding on its nature requires addressing the selection function of this sample.  }
 
 The two main branches  found in the tree might be attributed to the two AMRs which trace two different stellar populations. Perhaps the red group here is in fact the tail of the thick disk, since the red group traces a faster chemical enrichment history, an older population and a hotter dynamics overall. The ages of the red group here and in \cite{Jackson_2021MNRAS.502...32J} are not comparable as already discussed above. It is however difficult to be certain about this interpretation when looking at the sequences of the panels outside the solar neighbourhood (Fig.~\ref{fig:cn_feh_maps_r}), in which one of the two sequences seem to disappear. It would be interesting to see the branching pattern of phylogenetic trees outside the solar neighbourhood, but that is subject of another study. 
 
As extensively discussed by N20,  two sequences of age and metallicity in the solar neighbourhood can be explained considering the {\it two-infall} model for chemical evolution \citep{1997ApJ...477..765C}. In such model, there are two main episodes of star formation in the disk, which are different from each other. In the {\it two-infall} model the first episode was fast,  quickly enriching the interstellar medium with metals and forming the stars now belonging to the thick disk. Later on, a second episode which lasts until today, has been forming the thin disk stars. Both episodes are driven by some infall of metal-poor gas from outside. In the thin disk, however, the enrichment of gas higher in the inner regions,  which translates into a negative metallicity gradient with Galactic radii but not necessarily two separate populations, unless migration of stars is considered. 

Alternatively, \cite{Haywood_2019A&A...625A.105H} explains two different main episodes of disk evolution with a quenching of star formation in the thick disk.  The rapid star formation in the thick disk might have driven strong stellar feedback which could have enriched the outer disk through galactic fountains  while exhausting the gas reservoir in the disk to make new stars. The creation of the bar, then, could have caused through the outer Linblad resonances turbulence in the outer disk ($R>6$ kpc) which helped the ISM to mix and form stars again  \citep[see also recent summary description of this scenario in][]{Katz_2021arXiv210202082K}. Other ways to bring new gas that contains some metals into the disk  can be by a merging event. While the stars from the merging event will likely be deposited in the halo, the gas can mix with the disk which can lead to a new episode of star formation in the disk \citep{2019ApJ...883L...5B}.   To further constrain the details of such scenarios, having a more accurate value of the ages that the [C/N] abundances are reflecting is essential.


\subsection{Age metallicity relation,  radial migration}

As more accurate and precise data is available for stars in the Galaxy, the AMR is still a fundamental tool to study the formation of the disk \citep{2018MNRAS.475.5487S}. It particularly helps to constrain the effects of radial migration. \cite{Johnson_2021arXiv210309838J} presents an interesting  recent discussion about the effect of the dual AMR. 
The chemical evolution models discussed by \cite{Johnson_2021arXiv210309838J} study the effect on the AMR using different star formation rates. In particular, a star formation rate which they call "Late-Burst" accounting for a later star formation burst produced by perhaps a galaxy merger,  can create a bimodal AMR, especially at inner Galactic radii. They were however unable to confirm the duality of the AMR when comparing with data of \cite{Feuillet_2019MNRAS.489.1742F} which is their benchmark sample for observations, but perhaps considering a subset of stars would allow for higher precision in the data.   

The dual AMR here can also be the product of radial migration caused by resonances in the disk \citep{2010ApJ...722..112M}. As shown in that work, moving perturbes such as the rotation of the bar or the spiral structure in the disk can change the motions of the stars, inducing radial migration inward and outwards in the disk. As long as two perturbes act simultaneously, it is possible to obtain a bimodality in the change of angular momentum of the stars, regardless of the details of the past history of the Milky Way and the pattern speed or strength of such perturbes.  This bimodality might cause an effect in the metallicity distribution of the stars. The dual AMR seen only in the solar neighbourhood panel of Fig.~\ref{fig:cn_feh_maps_r} could thus be a signature of a spiral-bar resonance overlap, since it does not need to have the same effect everywhere in the disk.  


\subsection{The Sun's birthplace}
It is interesting to comment further on the statement made by \cite{Haywood_2019A&A...625A.105H} about the solar birth place. From radial migration arguments it is expected that the Sun formed at inner Galactic radii because its metallicity is higher than the metallicity of the ISM in the solar neighbourhood 4.5 Gyr ago. They argue, however, that when considering the TW diagrams at different  Galactic radii such as those displayed by \cite{2015ApJ...808..132H}, the solar [$\alpha/$Fe] abundance ratios do not fall in the bulk of the abundance distributions of the stars in the inner regions, which have generally higher [$\alpha$/Fe] abundances, but rather with the bulk of stars in outer regions.  

In Fig.~\ref{fig:abundances1}, the solar abundances lie in the blue groups for all panels. If the red group in Fig.~\ref{fig:abundances1} corresponds to the tail of the inner disk population and the blue group correspond to the tail of the outer disk population, then it is possible that indeed the Sun migrated inwards from outer regions in the disk as commented by  \cite{Haywood_2019A&A...625A.105H}.

\section{Conclusion}

Inspired by the latest results of \cite{Nissen_2020A&A...640A..81N} about the two age metallicity sequences in the solar neighbourhood using high-precision analysis of solar-type stars, this work presented an alternative relation of [C/N] and metallicity  of RC stars to test if the two sequences are present in larger and independent datasets. In red giants, [C/N] abundances can have a direct dependency on the stellar mass, hence ages. The advantage to use [C/N] instead of ages arises in the fact that measuring precise C and N abundances is more plausible than deriving ages. As concluded by N20, the two sequences were never seen before because stellar ages are still too uncertain. 

For this work, about 18,000 red clump stars selected from the APOGEE DR16 catalogue with precise measurements of [Fe/H], C and N, in addition to Galactic heights below 800 pc and solar scaled $\alpha$ abundances.  The density maps of [C/N] versus [Fe/H] for different Galactic radii revealed that indeed in the solar neighbourhood two separate sequences of [C/N] and age are present. Inner and outer regions in the disk, however, show just one dominant group. 

By analysing other elemental abundance ratios and dynamical distributions of stars belonging to each sequence it was possible to see that the sequences are different. { One is composed by stars  that are more metal-rich and [C/N] rich, suggesting an old population.  The other is composed by stars that cover a wider range in both, metallicity and [C/N], reaching lower values and suggesting to contain younger stars. } The old and metal rich population was shown to be kinematically hotter with more eccentric orbits than the younger population. This is consistent with current expectations of  metal-rich and old stars formed at inner parts of the disk moving to the solar vicinity through radial migration or simply passing by due to their eccentric orbits.  A dual AMR might also be predicted by discrete episodes of star formation, which is predicted in some models.  

{  The [C/N]-[Fe/H] sequence that covers a wider range in [Fe/H] showed two bumps of stars at slight different [C/N] abundances (hence ages). This bump could be related to the group found in the phylogenetic tree by \citep{Jackson_2021MNRAS.502...32J},  which was attributed to the product of a star formation burst. To conclude on this hypothesis however, counting stars should take selection effects into account}.  

Attributing the two populations to one being the tail of the inner disk and the other one the tail of the outer disk, each experiencing different chemical enrichment history, the solar abundances match better the outer sequence than the inner sequence. This supports \cite{Haywood_2019A&A...625A.105H} claims about the Sun forming in outer regions migrating towards the inner regions, and not the other way around, as believed by most studies. 

Modern survey data enable us to find structure in the disk in several ways, allowing us to make progress in the understanding of how our home galaxy formed. This work shows the power of having high resolution spectra of large number of stars, which give us not only the information about metallicity and the classical [$\alpha/$Fe] ratio, but other abundances too. This increases our chances to find structure in the Milky Way. Here [C/N] was used as key alternative to age, and other abundances such as Mn, Al, Ni and Ca were used to evaluate that the formation histories of the structures could be different provided the systematic uncertainties in elemental abundance measurements are properly addressed.  The revolution of combining multidimensional chemodynamical information in order to reveal the shared history of the stars in our Milky Way is just starting. 

\begin{acknowledgements}
The author acknowledges Danielle de Brito Silva for vibrant discussions, as well as Thomas M\"adler, Payel Das and Poul Erik Nissen for important feedback of early versions of this paper.  Sven Buder is further acknowledged for promoting and making research about the Tinsley-Wallerstein diagram, generating a page on wikipedia with the details. The author finally warmly thanks the referee, for their careful and friendly report that improved this manuscript. 
Figures resulted from the studying and following the example figures of the book of \cite{astroMLText}, adapting the routines to this dataset and purposes.  Work funded by FONDECYT REGULAR 1200703 and FONDECYT Iniciaci\'on 11170174.

\end{acknowledgements}

\bibliography{references}{}

\begin{thebibliography}{}
\expandafter\ifx\csname natexlab\endcsname\relax\def\natexlab#1{#1}\fi
\providecommand{\url}[1]{\href{#1}{#1}}
\providecommand{\dodoi}[1]{doi:~\href{http://doi.org/#1}{\nolinkurl{#1}}}
\providecommand{\doeprint}[1]{\href{http://ascl.net/#1}{\nolinkurl{http://ascl.net/#1}}}
\providecommand{\doarXiv}[1]{\href{https://arxiv.org/abs/#1}{\nolinkurl{https://arxiv.org/abs/#1}}}

\bibitem[{{Adibekyan} {et~al.}(2012){Adibekyan}, {Sousa}, {Santos}, {Delgado
  Mena}, {Gonz{\'a}lez Hern{\'a}ndez}, {Israelian}, {Mayor}, \&
  {Khachatryan}}]{Adibekyan_2012A&A...545A..32A}
{Adibekyan}, V.~Z., {Sousa}, S.~G., {Santos}, N.~C., {et~al.} 2012, \aap, 545,
  A32, \dodoi{10.1051/0004-6361/201219401}

\bibitem[{{Ahumada} {et~al.}(2020){Ahumada}, {Prieto}, {Almeida}, {Anders},
  {Anderson}, {Andrews}, {Anguiano}, {Arcodia}, {Armengaud}, {Aubert}, {Avila},
  {Avila-Reese}, {Badenes}, {Balland}, {Barger}, {Barrera-Ballesteros}, {Basu},
  {Bautista}, {Beaton}, {Beers}, {Benavides}, {Bender}, {Bernardi}, {Bershady},
  {Beutler}, {Bidin}, {Bird}, {Bizyaev}, {Blanc}, {Blanton}, {Boquien},
  {Borissova}, {Bovy}, {Brandt}, {Brinkmann}, {Brownstein}, {Bundy}, {Bureau},
  {Burgasser}, {Burtin}, {Cano-D{\'\i}az}, {Capasso}, {Cappellari}, {Carrera},
  {Chabanier}, {Chaplin}, {Chapman}, {Cherinka}, {Chiappini}, {Doohyun Choi},
  {Chojnowski}, {Chung}, {Clerc}, {Coffey}, {Comerford}, {Comparat}, {da
  Costa}, {Cousinou}, {Covey}, {Crane}, {Cunha}, {Ilha}, {Dai}, {Damsted},
  {Darling}, {Davidson}, {Davies}, {Dawson}, {De}, {de la Macorra}, {De Lee},
  {Queiroz}, {Deconto Machado}, {de la Torre}, {Dell'Agli}, {du Mas des
  Bourboux}, {Diamond-Stanic}, {Dillon}, {Donor}, {Drory}, {Duckworth},
  {Dwelly}, {Ebelke}, {Eftekharzadeh}, {Davis Eigenbrot}, {Elsworth},
  {Eracleous}, {Erfanianfar}, {Escoffier}, {Fan}, {Farr},
  {Fern{\'a}ndez-Trincado}, {Feuillet}, {Finoguenov}, {Fofie},
  {Fraser-McKelvie}, {Frinchaboy}, {Fromenteau}, {Fu}, {Galbany}, {Garcia},
  {Garc{\'\i}a-Hern{\'a}ndez}, {Oehmichen}, {Ge}, {Maia}, {Geisler}, {Gelfand},
  {Goddy}, {Gonzalez-Perez}, {Grabowski}, {Green}, {Grier}, {Guo}, {Guy},
  {Harding}, {Hasselquist}, {Hawken}, {Hayes}, {Hearty}, {Hekker}, {Hogg},
  {Holtzman}, {Horta}, {Hou}, {Hsieh}, {Huber}, {Hunt}, {Chitham}, {Imig},
  {Jaber}, {Angel}, {Johnson}, {Jones}, {J{\"o}nsson}, {Jullo}, {Kim},
  {Kinemuchi}, {Kirkpatrick}, {Kite}, {Klaene}, {Kneib}, {Kollmeier}, {Kong},
  {Kounkel}, {Krishnarao}, {Lacerna}, {Lan}, {Lane}, {Law}, {Le Goff}, {Leung},
  {Lewis}, {Li}, {Lian}, {Lin}, {Long}, {Longa-Pe{\~n}a}, {Lundgren}, {Lyke},
  {Ted Mackereth}, {MacLeod}, {Majewski}, {Manchado}, {Maraston}, {Martini},
  {Masseron}, {Masters}, {Mathur}, {McDermid}, {Merloni}, {Merrifield},
  {M{\'e}sz{\'a}ros}, {Miglio}, {Minniti}, {Minsley}, {Miyaji}, {Mohammad},
  {Mosser}, {Mueller}, {Muna}, {Mu{\~n}oz-Guti{\'e}rrez}, {Myers}, {Nadathur},
  {Nair}, {Nandra}, {do Nascimento}, {Nevin}, {Newman}, {Nidever}, {Nitschelm},
  {Noterdaeme}, {O'Connell}, {Olmstead}, {Oravetz}, {Oravetz}, {Osorio},
  {Pace}, {Padilla}, {Palanque-Delabrouille}, {Palicio}, {Pan}, {Pan},
  {Parker}, {Paviot}, {Peirani}, {Ram{\'r}ez}, {Penny}, {Percival},
  {Perez-Fournon}, {P{\'e}rez-R{\`a}fols}, {Petitjean}, {Pieri},
  {Pinsonneault}, {Poovelil}, {Povick}, {Prakash}, {Price-Whelan}, {Raddick},
  {Raichoor}, {Ray}, {Rembold}, {Rezaie}, {Riffel}, {Riffel}, {Rix}, {Robin},
  {Roman-Lopes}, {Rom{\'a}n-Z{\'u}{\~n}iga}, {Rose}, {Ross}, {Rossi},
  {Rowlands}, {Rubin}, {Salvato}, {S{\'a}nchez}, {S{\'a}nchez-Menguiano},
  {S{\'a}nchez-Gallego}, {Sayres}, {Schaefer}, {Schiavon}, {Schimoia},
  {Schlafly}, {Schlegel}, {Schneider}, {Schultheis}, {Schwope}, {Seo},
  {Serenelli}, {Shafieloo}, {Shamsi}, {Shao}, {Shen}, {Shetrone}, {Shirley},
  {Aguirre}, {Simon}, {Skrutskie}, {Slosar}, {Smethurst}, {Sobeck}, {Sodi},
  {Souto}, {Stark}, {Stassun}, {Steinmetz}, {Stello}, {Stermer},
  {Storchi-Bergmann}, {Streblyanska}, {Stringfellow}, {Stutz}, {Su{\'a}rez},
  {Sun}, {Taghizadeh-Popp}, {Talbot}, {Tayar}, {Thakar}, {Theriault}, {Thomas},
  {Thomas}, {Tinker}, {Tojeiro}, {Toledo}, {Tremonti}, {Troup}, {Tuttle},
  {Unda-Sanzana}, {Valentini}, {Vargas-Gonz{\'a}lez}, {Vargas-Maga{\~n}a},
  {V{\'a}zquez-Mata}, {Vivek}, {Wake}, {Wang}, {Weaver}, {Weijmans}, {Wild},
  {Wilson}, {Wilson}, {Wolthuis}, {Wood-Vasey}, {Yan}, {Yang}, {Y{\`e}che},
  {Zamora}, {Zarrouk}, {Zasowski}, {Zhang}, {Zhao}, {Zhao}, {Zheng}, {Zheng},
  {Zhu}, \& {Zou}}]{2020ApJS..249....3A}
{Ahumada}, R., {Prieto}, C.~A., {Almeida}, A., {et~al.} 2020, \apjs, 249, 3,
  \dodoi{10.3847/1538-4365/ab929e}

\bibitem[{{Allende Prieto} {et~al.}(2006){Allende Prieto}, {Beers}, {Wilhelm},
  {Newberg}, {Rockosi}, {Yanny}, \& {Lee}}]{2006ApJ...636..804A}
{Allende Prieto}, C., {Beers}, T.~C., {Wilhelm}, R., {et~al.} 2006, \apj, 636,
  804, \dodoi{10.1086/498131}

\bibitem[{{Amarsi} {et~al.}(2020{\natexlab{a}}){Amarsi}, {Grevesse}, {Grumer},
  {Asplund}, {Barklem}, \& {Collet}}]{2020A&A...636A.120A}
{Amarsi}, A.~M., {Grevesse}, N., {Grumer}, J., {et~al.} 2020{\natexlab{a}},
  \aap, 636, A120, \dodoi{10.1051/0004-6361/202037890}

\bibitem[{{Amarsi} {et~al.}(2019){Amarsi}, {Nissen}, \&
  {Sk{\'u}lad{\'o}ttir}}]{2019A&A...630A.104A}
{Amarsi}, A.~M., {Nissen}, P.~E., \& {Sk{\'u}lad{\'o}ttir}, {\'A}. 2019, \aap,
  630, A104, \dodoi{10.1051/0004-6361/201936265}

\bibitem[{{Amarsi} {et~al.}(2020{\natexlab{b}}){Amarsi}, {Lind}, {Osorio},
  {Nordlander}, {Bergemann}, {Reggiani}, {Wang}, {Buder}, {Asplund}, {Barklem},
  {Wehrhahn}, {Sk{\'u}lad{\'o}ttir}, {Kobayashi}, {Karakas}, {Gao},
  {Bland-Hawthorn}, {de Silva}, {Kos}, {Lewis}, {Martell}, {Sharma}, {Simpson},
  {Zucker}, {{\v{C}}otar}, {Horner}, \& {Galah
  Collaboration}}]{2020A&A...642A..62A}
{Amarsi}, A.~M., {Lind}, K., {Osorio}, Y., {et~al.} 2020{\natexlab{b}}, \aap,
  642, A62, \dodoi{10.1051/0004-6361/202038650}

\bibitem[{{Bensby} {et~al.}(2014){Bensby}, {Feltzing}, \&
  {Oey}}]{Bensby_2014A&A...562A..71B}
{Bensby}, T., {Feltzing}, S., \& {Oey}, M.~S. 2014, \aap, 562, A71,
  \dodoi{10.1051/0004-6361/201322631}

\bibitem[{{Bignone} {et~al.}(2019){Bignone}, {Helmi}, \&
  {Tissera}}]{2019ApJ...883L...5B}
{Bignone}, L.~A., {Helmi}, A., \& {Tissera}, P.~B. 2019, \apjl, 883, L5,
  \dodoi{10.3847/2041-8213/ab3e0e}

\bibitem[{{Bird} {et~al.}(2021){Bird}, {Loebman}, {Weinberg}, {Brooks},
  {Quinn}, \& {Christensen}}]{2021MNRAS.503.1815B}
{Bird}, J.~C., {Loebman}, S.~R., {Weinberg}, D.~H., {et~al.} 2021, \mnras, 503,
  1815, \dodoi{10.1093/mnras/stab289}

\bibitem[{{Blancato} {et~al.}(2019){Blancato}, {Ness}, {Johnston}, {Rybizki},
  \& {Bedell}}]{2019ApJ...883...34B}
{Blancato}, K., {Ness}, M., {Johnston}, K.~V., {Rybizki}, J., \& {Bedell}, M.
  2019, \apj, 883, 34, \dodoi{10.3847/1538-4357/ab39e5}

\bibitem[{{Bovy} {et~al.}(2014){Bovy}, {Nidever}, \& {et
  al}}]{Bovy_2014ApJ...790..127B}
{Bovy}, J., {Nidever}, D.~L., \& {et al}. 2014, \apj, 790, 127,
  \dodoi{10.1088/0004-637X/790/2/127}

\bibitem[{{Buder} {et~al.}(2020){Buder}, {Sharma}, {Kos}, {Amarsi},
  {Nordlander}, {Lind}, {Martell}, {Asplund}, {Bland-Hawthorn}, {Casey}, {De
  Silva}, {D'Orazi}, {Freeman}, {Hayden}, {Lewis}, {Lin}, {Schlesinger},
  {Simpson}, {Stello}, {Zucker}, {Zwitter}, {Beeson}, {Buck}, {Casagrande},
  {Clark}, {Cotar}, {Da Costa}, {de Grijs}, {Feuillet}, {Horner}, {Khanna},
  {Kafle}, {Liu}, {Montet}, {Nandakumar}, {Nataf}, {Ness}, {Spina}, {Traven},
  {Tepper-Garcia}, {Ting}, {Vogrincic}, {Wittenmyer}, {Zerjal}, \& {the GALAH
  collaboration}}]{Buder_2020arXiv201102505B}
{Buder}, S., {Sharma}, S., {Kos}, J., {et~al.} 2020, arXiv e-prints,
  arXiv:2011.02505.
\newblock \doarXiv{2011.02505}

\bibitem[{{Casagrande} {et~al.}(2011){Casagrande}, {Sch{\"o}nrich}, {Asplund},
  {Cassisi}, {Ram{\'\i}rez}, {Mel{\'e}ndez}, {Bensby}, \&
  {Feltzing}}]{2011A&A...530A.138C}
{Casagrande}, L., {Sch{\"o}nrich}, R., {Asplund}, M., {et~al.} 2011, \aap, 530,
  A138, \dodoi{10.1051/0004-6361/201016276}

\bibitem[{{Casali} {et~al.}(2019){Casali}, {Magrini}, {Tognelli}, {Jackson},
  {Jeffries}, {Lagarde}, {Tautvai{\v{s}}ien{\.{e}}}, {Masseron},
  {Degl'Innocenti}, {Prada Moroni}, {Kordopatis}, {Pancino}, {Randich},
  {Feltzing}, {Sahlholdt}, {Spina}, {Friel}, {Roccatagliata}, {Sanna},
  {Bragaglia}, {Drazdauskas}, {Mikolaitis}, {Minkevi{\v{c}}i{\={u}}t{\.{e}}},
  {Stonkut{\.{e}}}, {Chorniy}, {Bagdonas}, {Jimenez-Esteban}, {Martell}, {Van
  der Swaelmen}, {Gilmore}, {Vallenari}, {Bensby}, {Koposov}, {Korn}, {Worley},
  {Smiljanic}, {Bergemann}, {Carraro}, {Damiani}, {Prisinzano}, {Bonito},
  {Franciosini}, {Gonneau}, {Hourihane}, {Jofre}, {Lewis}, {Morbidelli},
  {Sacco}, {Sousa}, {Zaggia}, {Lanzafame}, {Heiter}, {Frasca}, \&
  {Bayo}}]{2019A&A...629A..62C}
{Casali}, G., {Magrini}, L., {Tognelli}, E., {et~al.} 2019, \aap, 629, A62,
  \dodoi{10.1051/0004-6361/201935282}

\bibitem[{{Casali} {et~al.}(2020){Casali}, {Spina}, {Magrini}, {Karakas},
  {Kobayashi}, {Casey}, {Feltzing}, {Van der Swaelmen}, {Tsantaki},
  {Jofr{\'e}}, {Bragaglia}, {Feuillet}, {Bensby}, {Biazzo}, {Gonneau},
  {Tautvai{\v{s}}ien{\.{e}}}, {Baratella}, {Roccatagliata}, {Pancino}, {Sousa},
  {Adibekyan}, {Martell}, {Bayo}, {Jackson}, {Jeffries}, {Gilmore}, {Randich},
  {Alfaro}, {Koposov}, {Korn}, {Recio-Blanco}, {Smiljanic}, {Franciosini},
  {Hourihane}, {Monaco}, {Morbidelli}, {Sacco}, {Worley}, \&
  {Zaggia}}]{Casali_2020A&A...639A.127C}
{Casali}, G., {Spina}, L., {Magrini}, L., {et~al.} 2020, \aap, 639, A127,
  \dodoi{10.1051/0004-6361/202038055}

\bibitem[{{Casamiquela} {et~al.}(2021){Casamiquela}, {Soubiran}, {Jofr{\'e}},
  {Chiappini}, {Lagarde}, {Tarricq}, {Carrera}, {Jordi},
  {Balaguer-N{\'u}{\~n}ez}, {Carbajo-Hijarrubia}, \&
  {Blanco-Cuaresma}}]{2021arXiv210314692C}
{Casamiquela}, L., {Soubiran}, C., {Jofr{\'e}}, P., {et~al.} 2021, arXiv
  e-prints, arXiv:2103.14692.
\newblock \doarXiv{2103.14692}

\bibitem[{{Chiappini} {et~al.}(1997){Chiappini}, {Matteucci}, \&
  {Gratton}}]{1997ApJ...477..765C}
{Chiappini}, C., {Matteucci}, F., \& {Gratton}, R. 1997, \apj, 477, 765,
  \dodoi{10.1086/303726}

\bibitem[{{Das} {et~al.}(2020){Das}, {Hawkins}, \&
  {Jofr{\'e}}}]{Das_2020MNRAS.493.5195D}
{Das}, P., {Hawkins}, K., \& {Jofr{\'e}}, P. 2020, \mnras, 493, 5195,
  \dodoi{10.1093/mnras/stz3537}

\bibitem[{{Das} \& {Sanders}(2019)}]{2019MNRAS.484..294D}
{Das}, P., \& {Sanders}, J.~L. 2019, \mnras, 484, 294,
  \dodoi{10.1093/mnras/sty2776}

\bibitem[{{Delgado Mena} {et~al.}(2019){Delgado Mena}, {Moya}, {Adibekyan},
  {Tsantaki}, {Gonz{\'a}lez Hern{\'a}ndez}, {Israelian}, {Davies}, {Chaplin},
  {Sousa}, {Ferreira}, \& {Santos}}]{delgado_2019A&A...624A..78D}
{Delgado Mena}, E., {Moya}, A., {Adibekyan}, V., {et~al.} 2019, \aap, 624, A78,
  \dodoi{10.1051/0004-6361/201834783}

\bibitem[{{Edvardsson} {et~al.}(1993){Edvardsson}, {Andersen}, {Gustafsson},
  {Lambert}, {Nissen}, \& {Tomkin}}]{1993A&A...275..101E}
{Edvardsson}, B., {Andersen}, J., {Gustafsson}, B., {et~al.} 1993, \aap, 500,
  391

\bibitem[{{Feltzing} {et~al.}(2020){Feltzing}, {Bowers}, \&
  {Agertz}}]{2020MNRAS.493.1419F}
{Feltzing}, S., {Bowers}, J.~B., \& {Agertz}, O. 2020, \mnras, 493, 1419,
  \dodoi{10.1093/mnras/staa340}

\bibitem[{{Feuillet} {et~al.}(2019){Feuillet}, {Frankel}, {Lind}, {Frinchaboy},
  {Garc{\'\i}a-Hern{\'a}ndez}, {Lane}, {Nitschelm}, \&
  {Roman-Lopes}}]{Feuillet_2019MNRAS.489.1742F}
{Feuillet}, D.~K., {Frankel}, N., {Lind}, K., {et~al.} 2019, \mnras, 489, 1742,
  \dodoi{10.1093/mnras/stz2221}

\bibitem[{{Garc{\'\i}a P{\'e}rez} {et~al.}(2016){Garc{\'\i}a P{\'e}rez},
  {Allende Prieto}, {Holtzman}, {Shetrone}, {M{\'e}sz{\'a}ros}, {Bizyaev},
  {Carrera}, {Cunha}, {Garc{\'\i}a-Hern{\'a}ndez}, {Johnson}, {Majewski},
  {Nidever}, {Schiavon}, {Shane}, {Smith}, {Sobeck}, {Troup}, {Zamora},
  {Weinberg}, {Bovy}, {Eisenstein}, {Feuillet}, {Frinchaboy}, {Hayden},
  {Hearty}, {Nguyen}, {O'Connell}, {Pinsonneault}, {Wilson}, \&
  {Zasowski}}]{Garcia_2016AJ....151..144G}
{Garc{\'\i}a P{\'e}rez}, A.~E., {Allende Prieto}, C., {Holtzman}, J.~A.,
  {et~al.} 2016, \aj, 151, 144, \dodoi{10.3847/0004-6256/151/6/144}

\bibitem[{{Gilmore} {et~al.}(2012){Gilmore}, {Randich}, {Asplund}, {Binney},
  {Bonifacio}, {Drew}, {Feltzing}, {Ferguson}, {Jeffries}, {Micela},
  {Negueruela}, {Prusti}, {Rix}, {Vallenari}, {Alfaro}, {Allende-Prieto},
  {Babusiaux}, {Bensby}, {Blomme}, {Bragaglia}, {Flaccomio}, {Fran{\c{c}}ois},
  {Irwin}, {Koposov}, {Korn}, {Lanzafame}, {Pancino}, {Paunzen},
  {Recio-Blanco}, {Sacco}, {Smiljanic}, {Van Eck}, {Walton}, {Aden}, {Aerts},
  {Affer}, {Alcala}, {Altavilla}, {Alves}, {Antoja}, {Arenou}, {Argiroffi},
  {Asensio Ramos}, {Bailer-Jones}, {Balaguer-Nunez}, {Bayo}, {Barbuy},
  {Barisevicius}, {Barrado y Navascues}, {Battistini}, {Bellas Velidis},
  {Bellazzini}, {Belokurov}, {Bergemann}, {Bertelli}, {Biazzo}, {Bienayme},
  {Bland-Hawthorn}, {Boeche}, {Bonito}, {Boudreault}, {Bouvier}, {Brandao},
  {Brown}, {de Bruijne}, {Burleigh}, {Caballero}, {Caffau}, {Calura},
  {Capuzzo-Dolcetta}, {Caramazza}, {Carraro}, {Casagrande}, {Casewell},
  {Chapman}, {Chiappini}, {Chorniy}, {Christlieb}, {Cignoni}, {Cocozza},
  {Colless}, {Collet}, {Collins}, {Correnti}, {Covino}, {Crnojevic}, {Cropper},
  {Cunha}, {Damiani}, {David}, {Delgado}, {Duffau}, {Edvardsson}, {Eldridge},
  {Enke}, {Eriksson}, {Evans}, {Eyer}, {Famaey}, {Fellhauer}, {Ferreras},
  {Figueras}, {Fiorentino}, {Flynn}, {Folha}, {Franciosini}, {Frasca},
  {Freeman}, {Fremat}, {Friel}, {Gaensicke}, {Gameiro}, {Garzon}, {Geier},
  {Geisler}, {Gerhard}, {Gibson}, {Gomboc}, {Gomez}, {Gonzalez-Fernandez},
  {Gonzalez Hernandez}, {Gosset}, {Grebel}, {Greimel}, {Groenewegen},
  {Grundahl}, {Guarcello}, {Gustafsson}, {Hadrava}, {Hatzidimitriou}, {Hambly},
  {Hammersley}, {Hansen}, {Haywood}, {Heber}, {Heiter}, {Held}, {Helmi},
  {Hensler}, {Herrero}, {Hill}, {Hodgkin}, {Huelamo}, {Huxor}, {Ibata},
  {Jackson}, {de Jong}, {Jonker}, {Jordan}, {Jordi}, {Jorissen}, {Katz},
  {Kawata}, {Keller}, {Kharchenko}, {Klement}, {Klutsch}, {Knude}, {Koch},
  {Kochukhov}, {Kontizas}, {Koubsky}, {Lallement}, {de Laverny}, {van Leeuwen},
  {Lemasle}, {Lewis}, {Lind}, {Lindstrom}, {Lobel}, {Lopez Santiago}, {Lucas},
  {Ludwig}, {Lueftinger}, {Magrini}, {Maiz Apellaniz}, {Maldonado}, {Marconi},
  {Marino}, {Martayan}, {Martinez-Valpuesta}, {Matijevic}, {McMahon},
  {Messina}, {Meyer}, {Miglio}, {Mikolaitis}, {Minchev}, {Minniti}, {Moitinho},
  {Momany}, {Monaco}, {Montalto}, {Monteiro}, {Monier}, {Montes}, {Mora},
  {Moraux}, {Morel}, {Mowlavi}, {Mucciarelli}, {Munari}, {Napiwotzki},
  {Nardetto}, {Naylor}, {Naze}, {Nelemans}, {Okamoto}, {Ortolani}, {Pace},
  {Palla}, {Palous}, {Parker}, {Penarrubia}, {Pillitteri}, {Piotto}, {Posbic},
  {Prisinzano}, {Puzeras}, {Quirrenbach}, {Ragaini}, {Read}, {Read}, {Reyle},
  {De Ridder}, {Robichon}, {Robin}, {Roeser}, {Romano}, {Royer}, {Ruchti},
  {Ruzicka}, {Ryan}, {Ryde}, {Santos}, {Sanz Forcada}, {Sarro Baro},
  {Sbordone}, {Schilbach}, {Schmeja}, {Schnurr}, {Schoenrich}, {Scholz},
  {Seabroke}, {Sharma}, {De Silva}, {Smith}, {Solano}, {Sordo}, {Soubiran},
  {Sousa}, {Spagna}, {Steffen}, {Steinmetz}, {Stelzer}, {Stempels},
  {Tabernero}, {Tautvaisiene}, {Thevenin}, {Torra}, {Tosi}, {Tolstoy}, {Turon},
  {Walker}, {Wambsganss}, {Worley}, {Venn}, {Vink}, {Wyse}, {Zaggia},
  {Zeilinger}, {Zoccali}, {Zorec}, {Zucker}, {Zwitter}, \& {Gaia-ESO Survey
  Team}}]{2012Msngr.147...25G}
{Gilmore}, G., {Randich}, S., {Asplund}, M., {et~al.} 2012, The Messenger, 147,
  25

\bibitem[{{Hasselquist} {et~al.}(2019){Hasselquist}, {Holtzman}, {Shetrone},
  {Tayar}, {Weinberg}, {Feuillet}, {Cunha}, {Pinsonneault}, {Johnson}, {Bird},
  {Beers}, {Schiavon}, {Minchev}, {Fern{\'a}ndez-Trincado},
  {Garc{\'\i}a-Hern{\'a}ndez}, {Nitschelm}, \&
  {Zamora}}]{Hasselquist_2019ApJ...871..181H}
{Hasselquist}, S., {Holtzman}, J.~A., {Shetrone}, M., {et~al.} 2019, \apj, 871,
  181, \dodoi{10.3847/1538-4357/aaf859}

\bibitem[{{Hawkins} {et~al.}(2015){Hawkins}, {Jofr{\'e}}, {Masseron}, \&
  {Gilmore}}]{Hawkins_2015MNRAS.453..758H}
{Hawkins}, K., {Jofr{\'e}}, P., {Masseron}, T., \& {Gilmore}, G. 2015, \mnras,
  453, 758, \dodoi{10.1093/mnras/stv1586}

\bibitem[{{Hayden} {et~al.}(2015){Hayden}, {Bovy}, {Holtzman}, {Nidever},
  {Bird}, {Weinberg}, {Andrews}, {Majewski}, {Allende Prieto}, {Anders},
  {Beers}, {Bizyaev}, {Chiappini}, {Cunha}, {Frinchaboy},
  {Garc{\'\i}a-Her{\'n}andez}, {Garc{\'\i}a P{\'e}rez}, {Girardi}, {Harding},
  {Hearty}, {Johnson}, {M{\'e}sz{\'a}ros}, {Minchev}, {O'Connell}, {Pan},
  {Robin}, {Schiavon}, {Schneider}, {Schultheis}, {Shetrone}, {Skrutskie},
  {Steinmetz}, {Smith}, {Wilson}, {Zamora}, \&
  {Zasowski}}]{2015ApJ...808..132H}
{Hayden}, M.~R., {Bovy}, J., {Holtzman}, J.~A., {et~al.} 2015, \apj, 808, 132,
  \dodoi{10.1088/0004-637X/808/2/132}

\bibitem[{{Hayden} {et~al.}(2020){Hayden}, {Sharma}, {Bland-Hawthorn}, {Spina},
  {Buder}, {Asplund}, {Casey}, {De Silva}, {D'Orazi}, {Freeman}, {Kos},
  {Lewis}, {Lin}, {Lind}, {Martell}, {Schlesinger}, {Simpson}, {Zucker},
  {Zwitter}, {Chen}, {Cotar}, {Feuillet}, {Horner}, {Joyce}, {Nordlander},
  {Stello}, {Tepper-Garcia}, {Ting}, {Wang}, \&
  {Wittenmyer}}]{Hayden_2020arXiv201113745H}
{Hayden}, M.~R., {Sharma}, S., {Bland-Hawthorn}, J., {et~al.} 2020, arXiv
  e-prints, arXiv:2011.13745.
\newblock \doarXiv{2011.13745}

\bibitem[{{Haywood} {et~al.}(2019){Haywood}, {Snaith}, {Lehnert}, {Di Matteo},
  \& {Khoperskov}}]{Haywood_2019A&A...625A.105H}
{Haywood}, M., {Snaith}, O., {Lehnert}, M.~D., {Di Matteo}, P., \&
  {Khoperskov}, S. 2019, \aap, 625, A105, \dodoi{10.1051/0004-6361/201834155}

\bibitem[{{Hekker} \& {Johnson}(2019)}]{2019MNRAS.487.4343H}
{Hekker}, S., \& {Johnson}, J.~A. 2019, \mnras, 487, 4343,
  \dodoi{10.1093/mnras/stz1554}

\bibitem[{{Horta} {et~al.}(2021){Horta}, {Schiavon}, {Mackereth}, {Pfeffer},
  {Mason}, {Kisku}, {Fragkoudi}, {Allende Prieto}, {Cunha}, {Hasselquist},
  {Holtzman}, {Majewski}, {Nataf}, {O'Connell}, {Schultheis}, \&
  {Smith}}]{Horta_2021MNRAS.500.1385H}
{Horta}, D., {Schiavon}, R.~P., {Mackereth}, J.~T., {et~al.} 2021, \mnras, 500,
  1385, \dodoi{10.1093/mnras/staa2987}

\bibitem[{{Huang} {et~al.}(2020){Huang}, {Sch{\"o}nrich}, {Zhang}, {Wu},
  {Chen}, {Wang}, {Xiang}, {Wang}, {Yuan}, {Li}, {Sun}, {Li}, \&
  {Liu}}]{2020ApJS..249...29H}
{Huang}, Y., {Sch{\"o}nrich}, R., {Zhang}, H., {et~al.} 2020, \apjs, 249, 29,
  \dodoi{10.3847/1538-4365/ab994f}

\bibitem[{{Iben}(1965)}]{1965ApJ...142.1447I}
{Iben}, Icko, J. 1965, \apj, 142, 1447, \dodoi{10.1086/148429}

\bibitem[{{Ivezi{\'c}} {et~al.}(2014){Ivezi{\'c}}, {Connolly}, {Vanderplas}, \&
  {Gray}}]{astroMLText}
{Ivezi{\'c}}, {\v Z}., {Connolly}, A., {Vanderplas}, J., \& {Gray}, A. 2014,
  Statistics, Data Mining and Machine Learning in Astronomy (Princeton
  University Press)

\bibitem[{{Izzard} {et~al.}(2018){Izzard}, {Preece}, {Jofre}, {Halabi},
  {Masseron}, \& {Tout}}]{2018MNRAS.473.2984I}
{Izzard}, R.~G., {Preece}, H., {Jofre}, P., {et~al.} 2018, \mnras, 473, 2984,
  \dodoi{10.1093/mnras/stx2355}

\bibitem[{{Jackson} {et~al.}(2021){Jackson}, {Jofr{\'e}}, {Yaxley}, {Das}, {de
  Brito Silva}, \& {Foley}}]{Jackson_2021MNRAS.502...32J}
{Jackson}, H., {Jofr{\'e}}, P., {Yaxley}, K., {et~al.} 2021, \mnras, 502, 32,
  \dodoi{10.1093/mnras/staa4028}

\bibitem[{{Jofr{\'e}} {et~al.}(2017{\natexlab{a}}){Jofr{\'e}}, {Das},
  {Bertranpetit}, \& {Foley}}]{Jofre_2017MNRAS.467.1140J}
{Jofr{\'e}}, P., {Das}, P., {Bertranpetit}, J., \& {Foley}, R.
  2017{\natexlab{a}}, \mnras, 467, 1140, \dodoi{10.1093/mnras/stx075}

\bibitem[{{Jofr{\'e}} {et~al.}(2017{\natexlab{b}}){Jofr{\'e}}, {Heiter}, \&
  {Buder}}]{2017ASInC..14...37J}
{Jofr{\'e}}, P., {Heiter}, U., \& {Buder}, S. 2017{\natexlab{b}}, in
  Astronomical Society of India Conference Series, Vol.~14, Astronomical
  Society of India Conference Series, 37--44.
\newblock \doarXiv{1709.09366}

\bibitem[{{Jofr{\'e}} {et~al.}(2019){Jofr{\'e}}, {Heiter}, \&
  {Soubiran}}]{Jofre_2019ARA&A..57..571J}
{Jofr{\'e}}, P., {Heiter}, U., \& {Soubiran}, C. 2019, \araa, 57, 571,
  \dodoi{10.1146/annurev-astro-091918-104509}

\bibitem[{{Jofr{\'e}} {et~al.}(2020){Jofr{\'e}}, {Jackson}, \& {Tucci
  Maia}}]{Jofre_2020A&A...633L...9J}
{Jofr{\'e}}, P., {Jackson}, H., \& {Tucci Maia}, M. 2020, \aap, 633, L9,
  \dodoi{10.1051/0004-6361/201937140}

\bibitem[{{Jofr{\'e}} {et~al.}(2015){Jofr{\'e}}, {M{\"a}dler}, {Gilmore},
  {Casey}, {Soubiran}, \& {Worley}}]{Jofre_2015MNRAS.453.1428J}
{Jofr{\'e}}, P., {M{\"a}dler}, T., {Gilmore}, G., {et~al.} 2015, \mnras, 453,
  1428, \dodoi{10.1093/mnras/stv1724}

\bibitem[{{Jofr{\'e}} {et~al.}(2016){Jofr{\'e}}, {Jorissen}, {Van Eck},
  {Izzard}, {Masseron}, {Hawkins}, {Gilmore}, {Paladini}, {Escorza},
  {Blanco-Cuaresma}, \& {Manick}}]{Jofre_2016A&A...595A..60J}
{Jofr{\'e}}, P., {Jorissen}, A., {Van Eck}, S., {et~al.} 2016, \aap, 595, A60,
  \dodoi{10.1051/0004-6361/201629356}

\bibitem[{{Jofr{\'e}} {et~al.}(2017{\natexlab{c}}){Jofr{\'e}}, {Traven},
  {Hawkins}, {Gilmore}, {Sanders}, {M{\"a}dler}, {Steinmetz}, {Kunder},
  {Kordopatis}, {McMillan}, {Bienaym{\'e}}, {Bland-Hawthorn}, {Gibson},
  {Grebel}, {Munari}, {Navarro}, {Parker}, {Reid}, {Seabroke}, \&
  {Zwitter}}]{Jofre_2017MNRAS.472.2517J}
{Jofr{\'e}}, P., {Traven}, G., {Hawkins}, K., {et~al.} 2017{\natexlab{c}},
  \mnras, 472, 2517, \dodoi{10.1093/mnras/stx1877}

\bibitem[{{Johnson} {et~al.}(2021){Johnson}, {Weinberg}, {Vincenzo}, {Bird},
  {Loebman}, {Brooks}, {Quinn}, {Christensen}, \&
  {Griffith}}]{Johnson_2021arXiv210309838J}
{Johnson}, J.~W., {Weinberg}, D.~H., {Vincenzo}, F., {et~al.} 2021, arXiv
  e-prints, arXiv:2103.09838.
\newblock \doarXiv{2103.09838}

\bibitem[{{J{\"o}nsson} {et~al.}(2020){J{\"o}nsson}, {Holtzman}, {Allende
  Prieto}, {Cunha}, {Garc{\'\i}a-Hern{\'a}ndez}, {Hasselquist}, {Masseron},
  {Osorio}, {Shetrone}, {Smith}, {Stringfellow}, {Bizyaev}, {Edvardsson},
  {Majewski}, {M{\'e}sz{\'a}ros}, {Souto}, {Zamora}, {Beaton}, {Bovy}, {Donor},
  {Pinsonneault}, {Poovelil}, \& {Sobeck}}]{Jonsson_2020AJ....160..120J}
{J{\"o}nsson}, H., {Holtzman}, J.~A., {Allende Prieto}, C., {et~al.} 2020, \aj,
  160, 120, \dodoi{10.3847/1538-3881/aba592}

\bibitem[{{Katz} {et~al.}(2021){Katz}, {Gomez}, {Haywood}, {Snaith}, \& {Di
  Matteo}}]{Katz_2021arXiv210202082K}
{Katz}, D., {Gomez}, A., {Haywood}, M., {Snaith}, O., \& {Di Matteo}, P. 2021,
  arXiv e-prints, arXiv:2102.02082.
\newblock \doarXiv{2102.02082}

\bibitem[{{Kobayashi} {et~al.}(2020{\natexlab{a}}){Kobayashi}, {Karakas}, \&
  {Lugaro}}]{Kobayashi_2020ApJ...900..179K}
{Kobayashi}, C., {Karakas}, A.~I., \& {Lugaro}, M. 2020{\natexlab{a}}, \apj,
  900, 179, \dodoi{10.3847/1538-4357/abae65}

\bibitem[{{Kobayashi} {et~al.}(2020{\natexlab{b}}){Kobayashi}, {Leung}, \&
  {Nomoto}}]{Kobayashi_2020ApJ...895..138K}
{Kobayashi}, C., {Leung}, S.-C., \& {Nomoto}, K. 2020{\natexlab{b}}, \apj, 895,
  138, \dodoi{10.3847/1538-4357/ab8e44}

\bibitem[{{Lagarde} {et~al.}(2017){Lagarde}, {Robin}, {Reyl{\'e}}, \&
  {Nasello}}]{Lagarde_2017A&A...601A..27L}
{Lagarde}, N., {Robin}, A.~C., {Reyl{\'e}}, C., \& {Nasello}, G. 2017, \aap,
  601, A27, \dodoi{10.1051/0004-6361/201630253}

\bibitem[{{Leung} \& {Bovy}(2019)}]{Leung_2019MNRAS.489.2079L}
{Leung}, H.~W., \& {Bovy}, J. 2019, \mnras, 489, 2079,
  \dodoi{10.1093/mnras/stz2245}

\bibitem[{{Mackereth} {et~al.}(2019){Mackereth}, {Bovy}, {Leung}, {Schiavon},
  {Trick}, {Chaplin}, {Cunha}, {Feuillet}, {Majewski}, {Martig}, {Miglio},
  {Nidever}, {Pinsonneault}, {Aguirre}, {Sobeck}, {Tayar}, \&
  {Zasowski}}]{Mackereth_2019MNRAS.489..176M}
{Mackereth}, J.~T., {Bovy}, J., {Leung}, H.~W., {et~al.} 2019, \mnras, 489,
  176, \dodoi{10.1093/mnras/stz1521}

\bibitem[{{Maiolino} \& {Mannucci}(2019)}]{2019A&ARv..27....3M}
{Maiolino}, R., \& {Mannucci}, F. 2019, \aapr, 27, 3,
  \dodoi{10.1007/s00159-018-0112-2}

\bibitem[{{Martig} {et~al.}(2016){Martig}, {Fouesneau}, {Rix}, {Ness},
  {M{\'e}sz{\'a}ros}, {Garc{\'\i}a-Hern{\'a}ndez}, {Pinsonneault}, {Serenelli},
  {Silva Aguirre}, \& {Zamora}}]{2016MNRAS.456.3655M}
{Martig}, M., {Fouesneau}, M., {Rix}, H.-W., {et~al.} 2016, \mnras, 456, 3655,
  \dodoi{10.1093/mnras/stv2830}

\bibitem[{{Masseron} \& {Gilmore}(2015)}]{2015MNRAS.453.1855M}
{Masseron}, T., \& {Gilmore}, G. 2015, \mnras, 453, 1855,
  \dodoi{10.1093/mnras/stv1731}

\bibitem[{{Masseron} \& {Hawkins}(2017)}]{2017A&A...597L...3M}
{Masseron}, T., \& {Hawkins}, K. 2017, \aap, 597, L3,
  \dodoi{10.1051/0004-6361/201629938}

\bibitem[{{Masseron} {et~al.}(2017){Masseron}, {Lagarde}, {Miglio}, {Elsworth},
  \& {Gilmore}}]{Masseron_2017MNRAS.464.3021M}
{Masseron}, T., {Lagarde}, N., {Miglio}, A., {Elsworth}, Y., \& {Gilmore}, G.
  2017, \mnras, 464, 3021, \dodoi{10.1093/mnras/stw2632}

\bibitem[{{Matteucci} \& {Greggio}(1986)}]{1986A&A...154..279M}
{Matteucci}, F., \& {Greggio}, L. 1986, \aap, 154, 279

\bibitem[{{Miglio} {et~al.}(2021){Miglio}, {Chiappini}, {Mackereth}, {Davies},
  {Brogaard}, {Casagrande}, {Chaplin}, {Girardi}, {Kawata}, {Khan}, {Izzard},
  {Montalb{\'a}n}, {Mosser}, {Vincenzo}, {Bossini}, {Noels}, {Rodrigues},
  {Valentini}, \& {Mandel}}]{Miglio_2021A&A...645A..85M}
{Miglio}, A., {Chiappini}, C., {Mackereth}, J.~T., {et~al.} 2021, \aap, 645,
  A85, \dodoi{10.1051/0004-6361/202038307}

\bibitem[{{Minchev} \& {Famaey}(2010)}]{2010ApJ...722..112M}
{Minchev}, I., \& {Famaey}, B. 2010, \apj, 722, 112,
  \dodoi{10.1088/0004-637X/722/1/112}

\bibitem[{{Ness} {et~al.}(2016){Ness}, {Hogg}, {Rix}, {Martig}, {Pinsonneault},
  \& {Ho}}]{2016ApJ...823..114N}
{Ness}, M., {Hogg}, D.~W., {Rix}, H.~W., {et~al.} 2016, \apj, 823, 114,
  \dodoi{10.3847/0004-637X/823/2/114}

\bibitem[{{Nissen}(2015)}]{Nissen_2015A&A...579A..52N}
{Nissen}, P.~E. 2015, \aap, 579, A52, \dodoi{10.1051/0004-6361/201526269}

\bibitem[{{Nissen} {et~al.}(2020){Nissen}, {Christensen-Dalsgaard},
  {Mosumgaard}, {Silva Aguirre}, {Spitoni}, \&
  {Verma}}]{Nissen_2020A&A...640A..81N}
{Nissen}, P.~E., {Christensen-Dalsgaard}, J., {Mosumgaard}, J.~R., {et~al.}
  2020, \aap, 640, A81, \dodoi{10.1051/0004-6361/202038300}

\bibitem[{{Nissen} \& {Gustafsson}(2018)}]{Nissen_2018A&ARv..26....6N}
{Nissen}, P.~E., \& {Gustafsson}, B. 2018, \aapr, 26, 6,
  \dodoi{10.1007/s00159-018-0111-3}

\bibitem[{{Palla}(2021)}]{2021MNRAS.503.3216P}
{Palla}, M. 2021, \mnras, 503, 3216, \dodoi{10.1093/mnras/stab293}

\bibitem[{{Pinsonneault} {et~al.}(2014){Pinsonneault}, {Elsworth}, {Epstein},
  {Hekker}, {M{\'e}sz{\'a}ros}, {Chaplin}, {Johnson}, {Garc{\'\i}a},
  {Holtzman}, {Mathur}, {Garc{\'\i}a P{\'e}rez}, {Silva Aguirre}, {Girardi},
  {Basu}, {Shetrone}, {Stello}, {Allende Prieto}, {An}, {Beck}, {Beers},
  {Bizyaev}, {Bloemen}, {Bovy}, {Cunha}, {De Ridder}, {Frinchaboy},
  {Garc{\'\i}a-Hern{\'a}ndez}, {Gilliland}, {Harding}, {Hearty}, {Huber},
  {Ivans}, {Kallinger}, {Majewski}, {Metcalfe}, {Miglio}, {Mosser}, {Muna},
  {Nidever}, {Schneider}, {Serenelli}, {Smith}, {Tayar}, {Zamora}, \&
  {Zasowski}}]{2014ApJS..215...19P}
{Pinsonneault}, M.~H., {Elsworth}, Y., {Epstein}, C., {et~al.} 2014, \apjs,
  215, 19, \dodoi{10.1088/0067-0049/215/2/19}

\bibitem[{{Randich} {et~al.}(2013){Randich}, {Gilmore}, \& {Gaia-ESO
  Consortium}}]{2013Msngr.154...47R}
{Randich}, S., {Gilmore}, G., \& {Gaia-ESO Consortium}. 2013, The Messenger,
  154, 47

\bibitem[{{Salaris} {et~al.}(2015){Salaris}, {Pietrinferni}, {Piersimoni}, \&
  {Cassisi}}]{2015A&A...583A..87S}
{Salaris}, M., {Pietrinferni}, A., {Piersimoni}, A.~M., \& {Cassisi}, S. 2015,
  \aap, 583, A87, \dodoi{10.1051/0004-6361/201526951}

\bibitem[{{Sharma} {et~al.}(2020){Sharma}, {Hayden}, {Bland-Hawthorn},
  {Stello}, {Buder}, {Zinn}, {Kallinger}, {Asplund}, {De Silva}, {Dorazi},
  {Freeman}, {Kos}, {Lewis}, {Lin}, {Lind}, {Martell}, {Simpson}, {Wittenmyer},
  {Zucker}, {Zwitter}, {Chen}, {Cotar}, {Esdaile}, {Hon}, {Horner}, {Huber},
  {Kafle}, {Khanna}, {Ting}, {Nataf}, {Nordlander}, {Saadon}, {Tinney},
  {Traven}, {Watson}, {Wright}, \& {Wyse}}]{2020arXiv200406556S}
{Sharma}, S., {Hayden}, M.~R., {Bland-Hawthorn}, J., {et~al.} 2020, arXiv
  e-prints, arXiv:2004.06556.
\newblock \doarXiv{2004.06556}

\bibitem[{{Silva Aguirre} {et~al.}(2018){Silva Aguirre}, {Bojsen-Hansen},
  {Slumstrup}, {Casagrande}, {Kawata}, {Ciuc{\v{a}}}, {Handberg}, {Lund},
  {Mosumgaard}, {Huber}, {Johnson}, {Pinsonneault}, {Serenelli}, {Stello},
  {Tayar}, {Bird}, {Cassisi}, {Hon}, {Martig}, {Nissen}, {Rix},
  {Sch{\"o}nrich}, {Sahlholdt}, {Trick}, \& {Yu}}]{2018MNRAS.475.5487S}
{Silva Aguirre}, V., {Bojsen-Hansen}, M., {Slumstrup}, D., {et~al.} 2018,
  \mnras, 475, 5487, \dodoi{10.1093/mnras/sty150}

\bibitem[{{Silva Aguirre} {et~al.}(2020){Silva Aguirre},
  {Christensen-Dalsgaard}, {Cassisi}, {Miller Bertolami}, {Serenelli},
  {Stello}, {Weiss}, {Angelou}, {Jiang}, {Lebreton}, {Spada}, {Bellinger},
  {Deheuvels}, {Ouazzani}, {Pietrinferni}, {Mosumgaard}, {Townsend}, {Battich},
  {Bossini}, {Constantino}, {Eggenberger}, {Hekker}, {Mazumdar}, {Miglio},
  {Nielsen}, \& {Salaris}}]{2020A&A...635A.164S}
{Silva Aguirre}, V., {Christensen-Dalsgaard}, J., {Cassisi}, S., {et~al.} 2020,
  \aap, 635, A164, \dodoi{10.1051/0004-6361/201935843}

\bibitem[{{Soubiran} {et~al.}(2008){Soubiran}, {Bienaym{\'e}}, {Mishenina}, \&
  {Kovtyukh}}]{2008A&A...480...91S}
{Soubiran}, C., {Bienaym{\'e}}, O., {Mishenina}, T.~V., \& {Kovtyukh}, V.~V.
  2008, \aap, 480, 91, \dodoi{10.1051/0004-6361:20078788}

\bibitem[{{Spina} {et~al.}(2018){Spina}, {Mel{\'e}ndez}, {Karakas}, {dos
  Santos}, {Bedell}, {Asplund}, {Ram{\'\i}rez}, {Yong}, {Alves-Brito}, {Bean},
  \& {Dreizler}}]{Spina_2018MNRAS.474.2580S}
{Spina}, L., {Mel{\'e}ndez}, J., {Karakas}, A.~I., {et~al.} 2018, \mnras, 474,
  2580, \dodoi{10.1093/mnras/stx2938}

\bibitem[{{Tayar} {et~al.}(2020){Tayar}, {Claytor}, {Huber}, \& {van
  Saders}}]{2020arXiv201207957T}
{Tayar}, J., {Claytor}, Z.~R., {Huber}, D., \& {van Saders}, J. 2020, arXiv
  e-prints, arXiv:2012.07957.
\newblock \doarXiv{2012.07957}

\bibitem[{{Tayar} {et~al.}(2017){Tayar}, {Somers}, {Pinsonneault}, {Stello},
  {Mints}, {Johnson}, {Zamora}, {Garc{\'\i}a-Hern{\'a}ndez}, {Maraston},
  {Serenelli}, {Allende Prieto}, {Bastien}, {Basu}, {Bird}, {Cohen}, {Cunha},
  {Elsworth}, {Garc{\'\i}a}, {Girardi}, {Hekker}, {Holtzman}, {Huber},
  {Mathur}, {M{\'e}sz{\'a}ros}, {Mosser}, {Shetrone}, {Silva Aguirre},
  {Stassun}, {Stringfellow}, {Zasowski}, \&
  {Roman-Lopes}}]{2017ApJ...840...17T}
{Tayar}, J., {Somers}, G., {Pinsonneault}, M.~H., {et~al.} 2017, \apj, 840, 17,
  \dodoi{10.3847/1538-4357/aa6a1e}

\bibitem[{{Tinsley}(1979)}]{1979ApJ...229.1046T}
{Tinsley}, B.~M. 1979, \apj, 229, 1046, \dodoi{10.1086/157039}

\bibitem[{{Xiang} {et~al.}(2017){Xiang}, {Liu}, {Shi}, {Yuan}, {Huang}, {Chen},
  {Wang}, {Tian}, {Wu}, {Yang}, {Zhang}, {Huo}, \& {Ren}}]{2017ApJS..232....2X}
{Xiang}, M., {Liu}, X., {Shi}, J., {et~al.} 2017, \apjs, 232, 2,
  \dodoi{10.3847/1538-4365/aa80e4}

\end{thebibliography}
\bibliographystyle{aasjournal}

\end{document}